\documentclass[twocolumn, times]{aastex63}

\usepackage{gensymb}
\usepackage{booktabs}
\usepackage{rotating}
\usepackage{float}
\usepackage{graphicx}
\usepackage{xkeyval}
\usepackage{ifpdf}
\usepackage{collectbox}

 \usepackage{color}
 \definecolor{purple}{rgb}{1,0,1}

\usepackage[yyyymmdd,hhmmss]{datetime}

\defcitealias{jess:2019}{Paper \scshape I}
\defcitealias{Dillon:2020}{Paper \scshape II}

\received{\today}
\revised{not yet}
\accepted{not yet}

\submitjournal{ApJ}
\shorttitle{Nanoflare Brightenings in Flare Stars}
\shortauthors{Grant et al.}

\begin{document}

   \title{Statistical Signatures of Nanoflare Activity. III. \\Evidence of Enhanced Nanoflaring Rates in Fully Convective stars as Observed by the NGTS}

\correspondingauthor{S.~D.~T. Grant}
\email{samuel.grant@qub.ac.uk}

\author[0000-0001-5170-9747]{S.~D.~T. Grant}
\affiliation{Astrophysics Research Centre, School of Mathematics and Physics, Queen’s University Belfast, Belfast, BT7 1NN, UK}

\author[0000-0002-9155-8039]{D.~B. Jess}
\affiliation{Astrophysics Research Centre, School of Mathematics and Physics, Queen’s University Belfast, Belfast, BT7 1NN, UK}
\affiliation{Department of Physics and Astronomy, California State University Northridge, Northridge, CA 91330, USA}

\author[0000-0003-2709-7693]{C.~J. Dillon}
\affiliation{Astrophysics Research Centre, School of Mathematics and Physics, Queen’s University Belfast, Belfast, BT7 1NN, UK}

\author[0000-0002-7725-6296]{M. Mathioudakis}
\affiliation{Astrophysics Research Centre, School of Mathematics and Physics, Queen’s University Belfast, Belfast, BT7 1NN, UK}

\author[0000-0002-9718-3266]{C.~A. Watson}
\affiliation{Astrophysics Research Centre, School of Mathematics and Physics, Queen’s University Belfast, Belfast, BT7 1NN, UK}

\author[0000-0003-0711-7992]{J.~A.~G. Jackman}
\affiliation{School of Earth and Space Exploration, Arizona State University, Tempe, AZ 85287, USA}
\affiliation{Department of Physics, University of Warwick, Gibbet Hill Road, Coventry CV4 7AL, UK}
\affiliation{Centre for Exoplanets and Habitability, University of Warwick, Gibbet Hill Road, Coventry CV4 7AL, UK}

\author[0000-0003-0343-7905]{D.~G. Jackson}
\affiliation{Astrophysics Research Centre, School of Mathematics and Physics, Queen’s University Belfast, Belfast, BT7 1NN, UK}

\author[0000-0003-1452-2240]{P.~J. Wheatley}
\affiliation{Department of Physics, University of Warwick, Gibbet Hill Road, Coventry CV4 7AL, UK}

\author{M.~R. Goad}
\affiliation{School of Physics and Astronomy, University of Leicester, University Road, Leicester, LE1 7RH, UK}

\author[0000-0003-2478-0120]{S.~L. Casewell}
\affiliation{School of Physics and Astronomy, University of Leicester, University Road, Leicester, LE1 7RH, UK}

\author[0000-0001-7416-7522]{D.~R. Anderson}
\affiliation{Department of Physics, University of Warwick, Gibbet Hill Road, Coventry CV4 7AL, UK}

\author[0000-0003-0684-7803]{M.~R. Burleigh}
\affiliation{School of Physics and Astronomy, University of Leicester, University Road, Leicester, LE1 7RH, UK}

\author[0000-0001-6604-5533]{R.~G. West}
\affiliation{Department of Physics, University of Warwick, Gibbet Hill Road, Coventry CV4 7AL, UK}

\author[0000-0002-1896-2377]{J.~I. Vines}
\affiliation{Departamento de Astronomia, Universidad de Chile, Casilla 36-D, Santiago, Chile}

\begin{abstract}
 \noindent Previous examinations of fully-convective M-dwarf stars have highlighted enhanced rates of nanoflare activity on these distant stellar sources. However, the specific role the convective boundary, which is believed to be present for spectral types earlier than M2.5V, plays on the observed nanoflare rates is not yet known. Here, we utilize a combination of statistical and Fourier techniques to examine M-dwarf stellar lightcurves that lie on either side of the convective boundary. We find that fully convective M2.5V (and later sub-types) stars have greatly enhanced nanoflare rates compared with their pre-dynamo mode transition counterparts. Specifically, we derive a flaring power-law index in the region of $3.00 \pm  0.20$, alongside a decay timescale of $200 \pm  100$~s for M2.5V and M3V stars, matching those seen in prior observations of similar stellar sub-types. Interestingly, M4V stars exhibit longer decay timescales of $450 \pm  50$~s, along with an increased power-law index of $3.10 \pm  0.18$, suggesting an interplay between the rate of nanoflare occurrence and the intrinsic plasma parameters, for example, the underlying Lundquist number. In contrast, partially convective (i.e., earlier sub-types from M0V to M2V) M-dwarf stars exhibit very weak nanoflare activity, which is not easily identifiable using statistical or Fourier techniques. This suggests that fully convective stellar atmospheres favor small-scale magnetic reconnection, leading to implications for the flare-energy budgets of these stars. Understanding why small-scale reconnection is enhanced in fully convective atmospheres may help solve questions relating to the dynamo behavior of these stellar sources.
\end{abstract}

\keywords{Computational methods (1965) --- Optical flares (1166) --- Stellar flares (1603) --- Flare stars (540)}
. 

\setcounter{table}{0}
\section{Introduction} 
\label{sec:intro}

Magnetic reconnection is a fundamental physical process in conducting plasmas that allows for the conversion of magnetic energy through the rearrangement of magnetic fields \citep[e.g.,][]{Priest:1986,Reale:2007,Cargill:2015a}. Magnetic reconnection came to initial prominence in astronomy as a proposed mechanism for observed solar flare activity. The first derived reconnection model was presented by \citet{Sweet1958} and \citet{Parker1957} as a two-dimensional magnetohydrodynamic (MHD) configuration, where there exists a long, thin diffusion region enabling magnetic fields to reconnect. The reconnection rate of this Sweet-Parker model was found to be inversely proportional to the square root of the dimensionless Lundquist number, $S$, which is defined as;
\begin{equation}
S = \frac{L v_{A}}{\eta} \ ,
\end{equation}
where $L$ is the length of the diffusion region, $v_{A}$ is the Alfv\'{e}n speed, and $\eta$ is the plasma resistivity \citep[a full derivation is presented by][]{Priest2007}. Predicted Sweet-Parker reconnection rates for solar conditions could not recreate actual observations and the derived energetics from Sweet-Parker events were orders-of-magnitude smaller than what was observed in the solar corona \citep[e.g.,][]{Crosby1993}. An alternative model was presented by \citet{Petschek1964}, which allowed for faster rates by permitting reconnection across far shorter length scales of the diffusion region \citep{Priest2007}. This remedies an issue with the Sweet-Parker model (reconnection rate $\propto 1/\sqrt{S}$) due to the Petschek reconnection rate being inversely proportional to the logarithm of the Lundquist number, which limits the influence of plasma conductivity and provides more robust similarities to the characteristics of large flare events \citep[e.g.,][]{Aschwanden2020}. 

Observationally, magnetic reconnection is characterised by an impulsive brightening as magnetic energy is converted into localised plasma heating, and is classically seen as stochastic, macroscopic events \citep[see the reviews of][]{Cargill:2004,Benz2010, Fletcher2011, Benz2017}. Subsequently, the plasma gradually cools over an extended period, which manifests as an exponentially decaying intensity from the time of maximum brightness \citep[e.g.,][]{Moffett1974, Moffett1976, Kowalski2013, Pitkin2014}. The quantification of this decay process is measured through the $e$-folding time, $\tau$, which is the time taken for the flare luminosity to decrease by a factor $1/e$. The magnitude of this value is dependent on the underlying local plasma conditions, such as the efficiencies of evaporative, non-evaporative, conductive, and radiative cooling processes \citep{Antiochos1978}.
Solar and stellar flare energies are governed by a power-law relationship \citep{Aschwanden:2000}. Here, the power-law exponent governs the frequency, $dN/dE$, of flaring events with an associated energy, $E$, through the relationship,  
\begin{equation}
\label{eqn:powerlaw}
    \frac{dN}{dE} ~\propto~ E^{-\alpha} \ ,
\end{equation}
where $\alpha$ represents the power-law index. The nature of a power-law relation dictates that low-energy flares will be many times more frequent than larger events, and that small-scale events become more energetically important as the power-law index, $\alpha$, increases. A range of power-law indices have been documented across varying solar and stellar flare energy windows, from $1.35 \leq \alpha \leq 2.90$ \citep{Berghmans:1998, Krucker:1998, Aschwanden:1999a, Parnell:2000, Benz:2002, Winebarger:2002, Aschwanden:2012, Aschwanden:2014, Aschwanden:2015}.

The stochastic nature of large solar flaring eventsindicates that they are too infrequent to be viable heating mechanism for the extra-ordinary temperatures of the solar corona, known as the coronal heating paradox. Instead, nanoflares, with individual energies around $10^{9}$ times less than their large-scale counterparts, were proposed as an alternative due to their higher occurrence rates \citep{Parker:1988}. In order to be considered as consequential to the flare energy budget, and thus atmospheric heating, it has been established that the minimum requirement is $\alpha \geq 2$ \citep{Parker:1988, Hudson:1991}. Due to their individual low energies, nanoflares are typically embedded within the noise floor of the measured intensity signals, leading to difficulties identifying individual nanoflare events. However, their higher occurrence frequency means they can be recovered from time series data using statistical techniques that do not rely on the individual identification of macroscopic intensity signals.

Building on the work of \citet{Terzo:2011} and \citet{Jess:2014}, 
\citet[][henceforth referred to as \citetalias{jess:2019}]{jess:2019} developed a robust method for nanoflare investigation. Through Monte-Carlo simulations, realistic nanoflare lightcurves were generated for a wide range of $\alpha$ and $\tau$ values, coupled with precise modeling of the noise characteristics of solar observables. Through this, \citetalias{jess:2019} was able to uncover nanoflare signatures in solar coronal observations, manifesting as asymmetric contributions to the intensity fluctuation distributions of coronal images, consistent with power-law distributions on the order of $1.82 \leq \alpha \leq 1.90$. Despite showing that the solar active region under study did not appear to contain the necessary nanoflare activity to influence coronal heating, \citetalias{jess:2019} provided a comprehensive method for analysing small-scale flare activity in intensity time series, and suggested the same techniques could be applied to stellar observations.

Subsequently, \citet[][henceforth referred to as \citetalias{Dillon:2020}]{Dillon:2020} utilised the techniques of \citetalias{jess:2019} on stellar lightcurves from A, K and M-type stars to investigate whether signals previously interpreted as $p$-mode oscillations in dMe flare stars could in-fact be caused by nanoflares. Fourier analysis of each spectral class showed no enhanced power around $p$-mode frequencies (i.e., ${1 - 1000}$~s) in A-types, power enhancement at $p$-mode frequencies in the K-type, and power enhancements across the entire frequency spectrum in M-dwarfs. These enhancements were classically seen as evidence of global wave activity generated in both the K and M-type stars, however Monte-Carlo simulations revealed that the M-type stars produced the asymmetric intensity fluctuation distribution effects consistent with nanoflares, as opposed to the symmetrical effects observed in the K-type distributions that are consistent with dominant oscillatory behaviour. The M-dwarf flare activity produced a power-law index of $\alpha = 3.25 \pm 0.20 $, greater than previously reported values, and enabled \citetalias{Dillon:2020} to show that the flaring rate was high enough for nanoflare signals to appear quasi-periodic when an entire stellar disk is integrated into a single lightcurve, thus explaining their influence on the resulting Fourier power spectra. The reason for the enhanced nanoflare power-law indices for M~dwarfs was not known, but it was theorized that the fully convective nature of these stars may be responsible.

While solar-like stars have a combination of convective and radiative zones bridging their core and visible surface, some stars operate in a fully convective manner. The change from partially- to fully-convective interiors has been related to the `convective boundary', distinguished by a lack of tachocline in later stars that are fully convective. The tachocline is a thin region of the stellar interior at the boundary between the radiative and convective zones that contains large radial shears due to the imbalance between the rigid radiative zone and the differentially rotating outer convective zone \citep{Spiegel:1992, Browning:2008}. \citet{Wright:2016} estimated that this transition occurs in M-dwarf stars around M3V and later, with recent studies suggesting a more precise transition at approximately M${2.1-2.3}$V \citep{Mullan:2020}. 

Convection is a primary driver of magnetic reconnection in stars \citep{Pederson:2016}. As magnetic reconnection is the driving force behind flares, changes to the convective nature of a star have important implications for the resulting flare dynamics. The tachocline is thought to play a role in strengthening magnetic fields in partially-convective stars such as the Sun, as the shear forces across the region can convert poloidal fields into stronger toroidal configurations \citep{Parfey:2007}. However, it is important to note that the tachocline is not necessary for magnetic field generation, since fully convective stars also exhibit magnetism, where the dynamo is theorized to be driven by helical turbulence \citep{Durney:1993, Browning:2008, Pipin:2009}, but this change in dynamo is still under debate. Indeed, \citet{Wright:2016} and \citet{Wright:2018} investigated the relationship between stellar rotation and activity levels for fully-convective late M-type dwarf stars. They found that the rotation/activity relationship for fully convective stars was almost indistinguishable from partially convective stars, suggesting the solar-type dynamo is independent of the presence of a tachocline. 

Returning to the results of \citet{Durney:1993}, \citet{Browning:2008} and \citet{Pipin:2009}, it is possible to hypothesize that the enhancement of nanoflaring rates is linked to the conversion of late M-type interiors to fully-convective, and hence to the consequent changes induced in the helical dynamo processes. Previous examinations of stellar flares on late-type MV stars have found a range of power-law indices. Early space-based observations produced values of $\alpha\sim1.5$ \citep{Collura1988}, although a range of partially- and fully-convective MV stars were included in the sample studied. Subsequent studies of late-type stars produced indices $\alpha>2$ and showed that there were no discrepancies between ground- and space-based observatories in these calculations \citep{Robinson:1995, Robinson:1999}. A trend developed where reported power-law indices increased as the complexity of the methods for isolating marginal flare signals developed \citep[e.g.,][]{Gudel:2003, gudel:2004, Welsh:2006, Hawley:2014}, with values as high as  $\alpha = 2.7$. However, this remains below the nanoflare power-law index ($\alpha \approx 3.25$) presented in \citetalias{Dillon:2020}. This may be a result of the novel detection techniques of \citetalias{jess:2019} providing unprecedented access to the lower energy spectrum of flares, as it has been seen in solar observations that the power-law index, $\alpha$, can change depending on the size and energetics of the flares under consideration in the sample \citep[e.g.,][]{Wang:2013, Ryan:2016, Milligan:2020}. Therefore, there may exist a similar noticable shift in the gradient of the power-law index around the transition between large and small-scale flares in MV stars. Hence, it is important to examine the power-law indices associated with nanoflare activity across a wide range of MV spectral sub-types with the techniques described in \citetalias{jess:2019} and \citetalias{Dillon:2020} to better understand the influence of fully-convective interiors in the generation of nanoflares.

As discussed in \citetalias{Dillon:2020}, an alternative source for the enhanced rate of small-scale reconnection may be the fully-convective stars having plasma with a higher resistivity value \citep{Mohanty:2002}, which lowers the associated plasma Lundquist numbers. Small-scale flaring has been shown to occur more favorably via Sweet-Parker reconnection  \citep{Tsuneta:2004},  thus enhanced nanoflaring can be expected in stars with low plasma Lundquist numbers. If nanoflare rates are enhanced in fully convective stars, then investigating whether this is due to the change in dynamo, or down to the plasma resistivity, could answer important questions regarding the dynamo physics in operation in these stars. 

\section{Background to Previous Statistical Stellar Nanoflare Analysis}
\label{sec:previousanalysis}

\begin{figure*}[!t]
\centering
\includegraphics[ clip=true, angle=0, width=\textwidth]{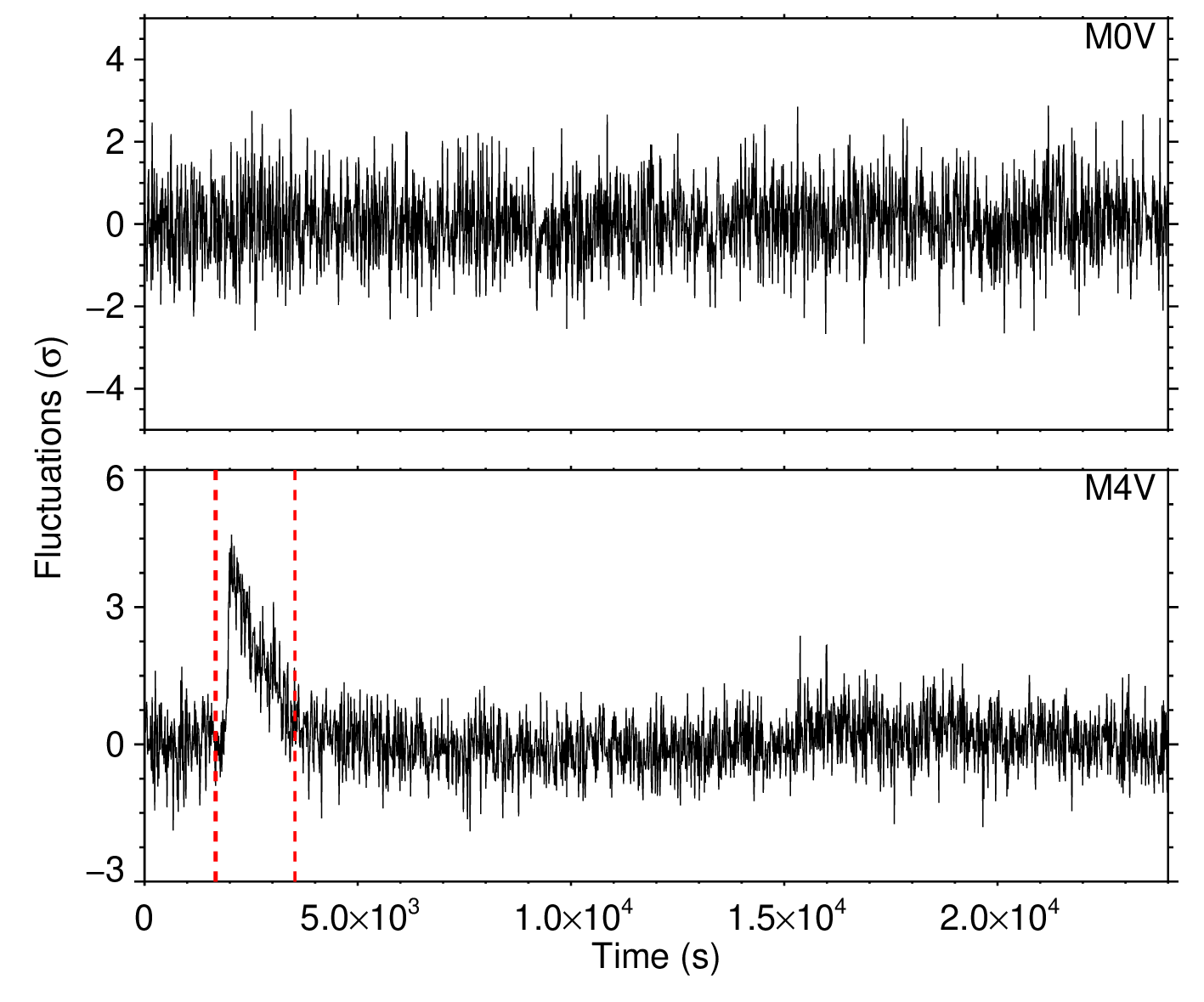}
\caption{Sample lightcurves of NGTS~J052346.3-361114 (M0V-type; top panel), and NGTS~J050423.8-373021 (M4V-type; lower panel) spanning 24{\,}000~s, each normalized by their respective standard deviations, $\sigma_{N}$. The region highlighted in red in the lower panel denotes the intensity values that are removed from consideration due to an excursion above 3$\sigma_{N}$, as is the formality for macroscopic flare signatures. }
\label{Lightcurves}
\end{figure*}

\citetalias{Dillon:2020} used a combination of statistical and Fourier analyses to investigate nanoflare populations in a variety of stellar sources. Through comparisons between observational lightcurves and synthetic time series with simulated nanoflare signals, enhanced nanoflare activity in late-type MV stars was observed, through a larger power-law index when the nanoflare occurrence rates were plotted as a function of their underlying energy. In order to determine why the power-law indices in these stars were significantly larger than in other stellar and solar studies, we re-apply the analysis techniques put forward by \citetalias{jess:2019} and \citetalias{Dillon:2020} to M-dwarf stars across a range of different spectral types. By determining the change in observed nanoflare properties (i.e., the change in the statistical and Fourier parameters associated with the embedded nanoflare conditions) either side of the fully-convective boundary, we aim to identify the role a fully-convective atmosphere plays in nanoflare occurrence rates. This work constitutes the third contribution to a series of studies on nanoflare behavior in solar and stellar atmospheres. Given the consistency of techniques applied across these studies, it is prudent to provide contextual information on previous works. For detailed discussions of the analysis procedures and modeling set-ups, it is advised to consult \citetalias{jess:2019} and \citetalias{Dillon:2020} directly.

The initial detection method for nanoflare signatures involves the statistical analyses of quiescent intensity fluctuations following a traditional Z-scores approach \citep{Sprinthall:1990}. A histogram of the fluctuations is generated, with two distinct statistical deviations away from a standardized Gaussian distribution providing evidence of nanoflares. These signatures, identified by \citet{Terzo:2011} and \citet{Jess:2014}, were diagnosed through the modeling of the observed lightcurves by embedding the rapid rises and exponential decays of intensities around the noise limit that are associated with the energetics of nanoflares. The first detection characteristic is a negative median offset of the intensity fluctuation distribution, whereby the median value of the histogram is $< 0~\sigma_N$, i.e., offset from the mean of the distribution that is equal to $0~\sigma_{N}$ following the application of polynomial detrending. This was shown to be associated with the exponentially decaying nature of the nanoflare lightcurve. The decay phase produces more significant negative fluctuations below the mean than the relatively brief elevated signal of the flare event, thus providing an offset between the median and mean that is directly measurable. The second signature of nanoflare activity is an excess of fluctuations at $\sim 2~\sigma_N$, which is caused by the associated energetics of nanoflares producing consistent peak brightenings around $\sim 2~\sigma_N$ (see \citealt{Jess:2014} and \citetalias{jess:2019} for further details on flare energy modeling). This gives rise to an asymmetric distribution with a slight excess of fluctuations visible at $\sim 2~\sigma_N$ in the corresponding histogram, which is best categorized through Fisher skewness coefficients. In addition to the primary nanoflare indicators described above, benchmarks on the shapes and widths of the intensity fluctuation distributions are provided through calculation of the kurtosis and $\zeta$ values, where $\zeta$ is the ratio of the full-width at eighth-maximum to that of the full-width at half-maximum (i.e., FW$\frac{1}{8}$M-to-FWHM ratio) of the resulting distribution. These parameters allow for additional information on the nature of the distribution, with kurtosis linked to the prevalence of outliers at high $\sigma_N$ values, and deviations from the established $\zeta$ value for a normal distribution ($\zeta=1.73$) revealing non-linear distributions of measurements around the mean.

As established in \citetalias{jess:2019} \& \citetalias{Dillon:2020}, a quiescent lightcurve exhibiting both of these statistical signals, identified through the median offset and Fisher skewness, contains embedded nanoflare signatures. \citet{Terzo:2011} presented an observation of nanoflare activity that only induced a median offset in the distribution, due to the weak nanoflare signal not producing sufficient peak amplitudes to influence the positive wing of the distribution. However, both effects presenting in observations, and corroborated by the four chief diagnostics, cannot be explained by any alternative mechanism. In an observation of purely ambient solar or stellar plasma, noise would be the only artifact. However, noise fluctuations follow a standard Gaussian distribution as a result of Poisson statistics tending to a Gaussian in the limit of large number statistics \citep{Terzo:2011}. Solar and stellar observations contain a variety of oscillatory phenomena, ranging from the signatures of MHD waves in their atmospheres \citep[see the review of][]{Jess2023} to modulation caused by the rotation of the star and starspots. In the context of this study, these signals have no influence over the asymmetric effects under consideration, as the sinusoidal nature of linear oscillations produces a symmetric distribution around the mean. Starspots, characterized by periodic reductions in intensity for times related to rotation, will decrease both the median and mean of the sample, providing a minimal effect in the offset between the two. Their effect on the timeseries is also mitigated through the application of polynomial detrending, as the periodicity of their effect is on the order of multiple days. MHD waves are also capable of steepening into non-linear shocks that produce notable intensity enhancements over a short period \citep[e.g.,][]{Carlsson1997, Grant2018}. However, these shocks display a `saw-tooth' pattern to their intensity morphology, in contrast to the exponential decay of flares, and thus would not reproduce the statistical effects described above. Therefore, with these effects discounted, and any macroscopic flare events robustly removed from the data (see Section~\ref{sec:observations}), there is a confidence that the statistical effects derived from the intensity fluctuation distributions are positive signatures of nanoflare activity.

Through inspection of nanoflare observations with insight from simulations, \citet{Terzo:2011} and \citet{Jess:2014} found the interval between nanoflares in a single lightcurve to be $\sim360$~s, a similar frequency to ubiquitous $p$-mode signatures \citep[$\sim 1-10$~mHz;][]{Andrews:1989, Rodriquez:2014,Rodriguez:2016}. Subsequently, \citetalias{jess:2019} were able to derive the frequency of flaring events given their power-law index, showing that the greater the power-law index, the higher the frequency of nanoflares in a given sample (see Figure~4 of \citetalias{jess:2019}). With the low flaring intervals and high frequencies found, it was suggested that nanoflare signals, when integrated across a field-of-view, can no longer be viewed as stochastic events like their macroscopic counterparts. Instead, they can be considered to be a quasi-periodic phenomena, particularly given the exceptionally large power law index of $\alpha \sim 3.25$ reported by \citetalias{Dillon:2020}.

The quasi-periodic nature of nanoflares was successfully utilized in \citetalias{Dillon:2020} to further verify their detection. When inspecting the statistical distributions of K- and M-type stars, there was the potential, but marginal, signature of nanoflares in K-types, as opposed to clear nanoflare signals in the M-types. \citetalias{Dillon:2020} subsequently employed Fourier techniques to distinguish the quasi-periodicities of stellar nanoflares. Power spectral densities (PSDs) were computed for the longest continuous time series common across all stars in order to maximize the resulting frequency resolution. These PSDs revealed power at $p$-mode frequencies in both types, however the nature of that power differed, with M-types showing power enhancement across a wide range of frequencies, synonymous with nanoflare activity from a range of interval times/frequencies, as opposed to the strictly oscillatory nature of K-type atmospheres (see Figure~3 of \citetalias{Dillon:2020}). Additionally, the PSDs in M-types displayed a prominent spectral slope following the peak energy value, which was also found to be due to the underlying nanoflare signal. Fourier analysis is therefore a valuable tool in both confirming nanoflare signal in the source, and differentiating between those atmospheres with wave and nanoflare interplay (i.e., partially-convective interiors) and where nanoflares dominate the energy landscape (i.e., fully-convective interiors).

Throughout these works, the observational findings have been substantiated and interpreted through simulated nanoflare time series. \citetalias{jess:2019} devised the methodology and analysis techniques, based on the Monte-Carlo modeling of flare intensities with additionally added camera-specific noise signatures, coupled with a range of typical nanoflare amplitudes and decay rates, and provides a detailed description of the set-up. \citetalias{Dillon:2020} took these techniques and tailored them for a strictly stellar scenario, including the remodeling of the noise profiles and resizing the observed area to mimic a full stellar disk, as opposed to a sub-section of the solar atmosphere used in \citetalias{jess:2019}. Flare energies between $10^{22} - 10^{25}$~erg, typical of nanoflares, were used alongside a linear scaling relationship to reproduce the observed counts. Each simulated lightcurve has two variables controlling the nanoflare input: the power-law index, $\alpha$, and the $e$-folding time, $\tau$. Lightcurves were then generated for a dense grid of parameter ranges, $1 \leq \alpha \leq 4$ (in steps of 0.05) \& $5 \leq \tau \leq 500$ (in steps of 5~s), consistent with previous viable observation ranges \citep{Terzo:2011, Jess:2014}. This synthesis produced 6100 lightcurves embedded with characteristic noise and unique nanoflare configurations. Subsequent comparison between synthetic and observed lightcurves is achieved through forming distributions of each synthetic time-series, and identifying the configuration with matching median offset, Fisher skewness, kurtosis and $\zeta$ values (see Figure~5 of \citetalias{jess:2019}). Given the stringent requirements of this matching criteria, unique solutions are found for the observables in both \citetalias{jess:2019} and \citetalias{Dillon:2020}. Given the computational demands, the simulations presented in \citetalias{Dillon:2020} are used in this study, thus further details on their reproducability can be found in the original paper. 

In this paper, we apply proven statistical nanoflare analysis techniques to a wide range of M-type stars that lie on either side of the predicted fully-convective boundary. We compare their statistical and Fourier properties to the simulations generated in \citetalias{Dillon:2020} in order to determine the probable underlying nanoflare conditions and the effect of the convective boundary on the uncovered nanoflare properties.

\section{Observations With NGTS} 
\label{sec:observations}

To remain consistent with \citetalias{Dillon:2020}, the Next Generation Transit Survey \citep[NGTS;][]{Wheatley:2017} was utilized to obtain the observations. The long time series (each in excess of $10^5$ frames) and short cadence ($\sim12$~s) available for thousands of M-type stars allow for the accumulation of suitable number statistics necessary for nanoflare analyses. The initial spectral classification generated by the NGTS pipeline \citep[which utilizes Spectral Energy Distribution fitting, see section 5.1.1 in][]{Wheatley:2017} was combined with stellar parameters from the TESS Input Catalog Version~8  \citep[TIC~V8;][]{Stassun:2018}, to ensure robust spectral sub-type identification.

\begin{figure*}[!t]
\centering
\includegraphics[ clip=true, angle=0, width=\textwidth]{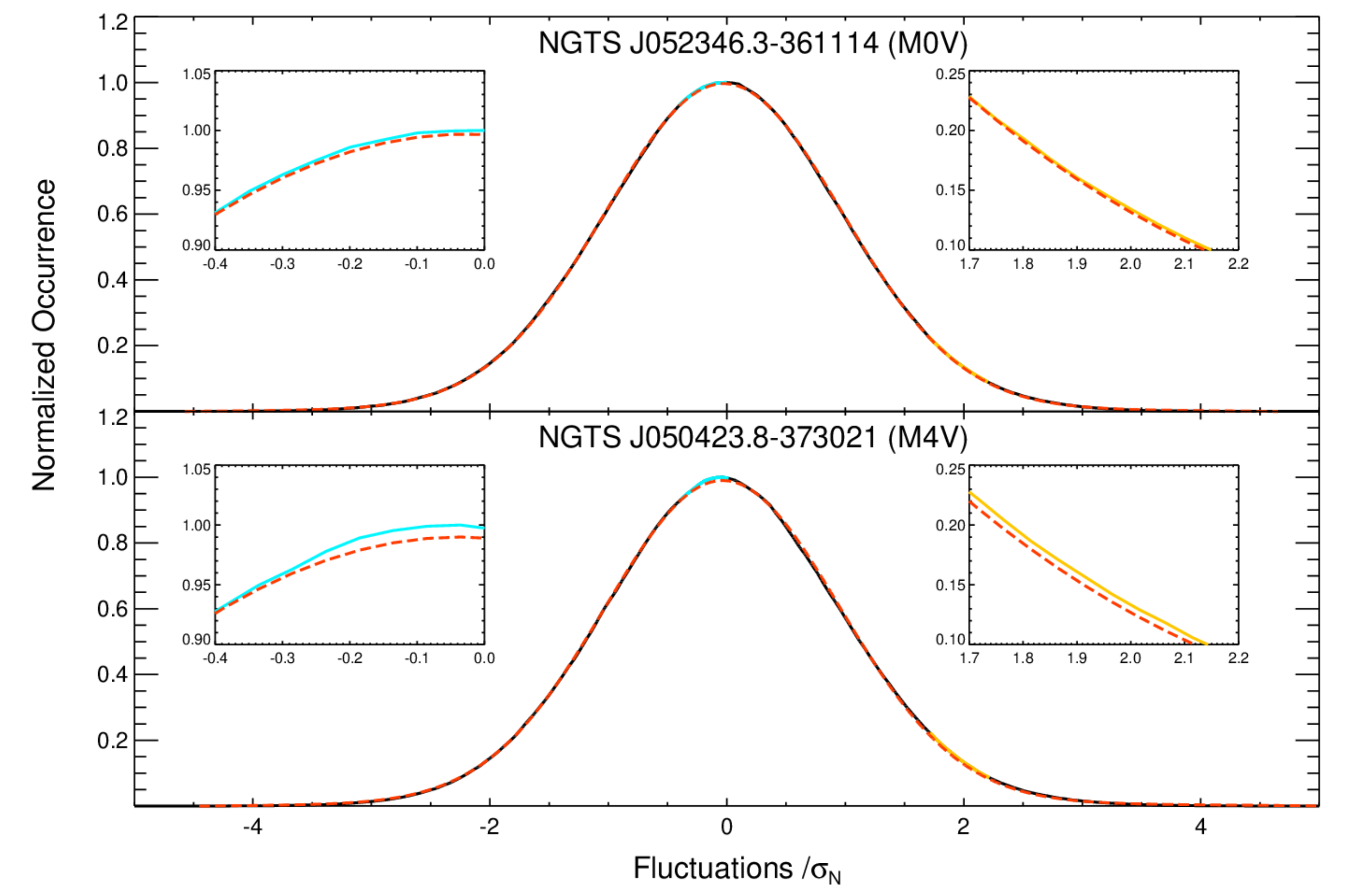}
\caption{Histograms of intensity fluctuations, each normalized by their respective standard deviations, $\sigma_{N}$, for the NGTS~J052346.3-361114 (M0V-type; top panel), and NGTS~J050423.8-373021 (M4V-type; lower panel) lightcurves. A standardized Gaussian profile is overplotted in each panel using a dashed red line for reference. The M4V-type distribution has a negative median offset with respect to the Gaussian, in addition to elevated occurrences at $\sim 2~\sigma_{N}$, which is consistent with the statistical signatures of nanoflare activity. On the other hand, the M0V-type intensity fluctuations provide effectively zero negative median offset, and no elevated occurrences at $\sim 2~\sigma_{N}$. This is inconsistent with clear statistical signatures of nanoflare activity, with the resulting distribution remaining more consistent with the presence of photon-based shot noise. Zoomed insets highlight the ranges spanning $-0.4 \le \sigma_{N} \le 0.0$ and $1.7 \le \sigma_{N} \le 2.2$, where negative median offsets and occurrence excesses, respectively, are clearly visible for the M4V stellar source. For improved clarity, the blue and gold lines display the corresponding distributions in each zoomed panel.}
\label{Histogram}
\end{figure*}

To ensure the ideal sample of objects for study, a number of selection criteria were applied to the catalog extracted from NGTS to remove unwanted artifacts. Initially, the magnitude of the stars were constrained to ensure similar photon noise characteristics for each object. Thus, only stars with magnitudes matching the range of the previous study of \citetalias{Dillon:2020} (spanning NGTS magnitudes of $\sim12-14$) were progressed. This ensured that the magnitude of the fluctuations in each spectral type were approximately equivalent, with M0V and M4V stars exhibiting standard deviations of $1.6 \pm 0.1$\% and $2.0 \pm 0.6$\%, respectively. The average standard deviation of the time-series across the full sample was $2.0 \pm 0.8$\% of the mean, thus an equivalent fluctuation profile can be applied to the modeling. Next, in a similar manner to \citet{Jackson2023}, complementary photometric and astrometric data from {\it{Gaia DR3}} \citep{Gaia2016, Gaia2022} were utilized to exclude unwanted candidates from the sample. In particular, astrometric excess noise analysis was applied to exclude any binary systems \citep{Evans2018}, and the photometric filtering processes of \citet{Arenou2018} identified any blended sources. A final step was to consider the rotation rate of the stars in the sample. Given the length of the NGTS time series, it is not possible to extract the periods associated with slow rotating M-dwarfs (i.e., above 30 days). Therefore, generalized Lomb-Scargle techniques \citep{Lomb1976, Scargle1982} were applied to the NGTS data to identify fast rotators and exclude them from the sample. This is done to ensure the sample contains similarly `slow' rotators, as defined in previous studies \citep[e.g.,][]{Mondrik2019}, thus removing the influence rotation rate has on increasing stellar activity \citep{West2008, Candelaresi2014}. Only two candidates exhibited definable periodicities and were thus excluded, an M2V \& an M4V with periods of 21.1~days and 19.8~days, respectively.

After accounting for magnitude considerations, avoiding blended sources, ensuring TIC matching and excluding fast rotators, we were able to find 5 stars for each spectral sub-type, consisting of M0V, M1V, M2V, M2.5V, M3V, and M4V. The stellar properties (NGTS identifier, RA/Dec, magnitude, etc.) of these candidates are provided in Table~{\ref{tab:StellarParam}}. Only one suitable M5V star with TIC-derived stellar parameters could be identified, and no sub-types later than this were found. The intrinsic brightness of M-dwarfs decreases with increasing sub-type \citep{Yang:2017}, leading to difficulty in identifying suitable candidate stars with the desired brightness properties. Future investigations of post-M4V stars may be fruitful, but identifying a suitable number of candidates may prove difficult with existing instrumentation. Hence, we limit our current study within the range of M0V -- M4V, where we have multiple candidates available for comparison. This range also overlaps well with the predicted dynamo mode transition to fully convective\citep[M${2.1-2.3}$V;][]{Mullan:2020}, making it suitable for the study of the role that fully-convective starsnplay in the resulting nanoflare activity.

The lightcurves were background corrected and flat-fielded via the NGTS data reduction pipeline described and visualized in \cite{Wheatley:2017}. This pipeline calculates a relative error in the flux at each data point in the time series. This error correlates with cloudy weather and/or high airmass values. Any fluctuations in this error exceeding 1$\sigma$ above the mean value were removed, resulting in $\sim$10\% of each time series being omitted. This removed any data that had statistically significant increases in its associated flux uncertainties, therefore preventing any large flux errors (largely due to poor seeing conditions) from contaminating the final time series.

\begin{deluxetable*}{ccccc}[!t]
\label{tab:Avg_stats}
\tablecaption{Averaged characteristics of the statistical properties by each spectral type.} 
\tablewidth{0pt}
\tablehead{
\colhead{Spectral type} & \colhead{Median offset ($\sigma_N$) }  & \colhead{Fisher skewness }  & \colhead{$\zeta$ ratio}&  \colhead{Kurtosis}   
}
\startdata
M0V &   $-0.040 \pm 0.008$ & $0.004 \pm 0.002$ & $1.740 \pm 0.005$ & $0.128 \pm 0.010$  \\
M1V &   $-0.040 \pm 0.008$ & $0.008 \pm 0.002$ & $1.746 \pm 0.005$ & $0.180 \pm 0.013$  \\
M2V &   $-0.030 \pm 0.006$ & $0.003 \pm 0.011$ & $1.766 \pm 0.010$ & $0.196 \pm 0.037$  \\
M2.5V&  $-0.050 \pm 0.000$ & $0.019 \pm 0.003$ & $1.739 \pm 0.004$ & $0.227 \pm 0.072$   \\
M3V &   $-0.050 \pm 0.000$ & $0.025 \pm 0.004$ & $1.750 \pm 0.006$ & $0.180 \pm 0.017$  \\
M4V &   $-0.050 \pm 0.000$ & $0.051 \pm 0.014$ & $1.754 \pm 0.010$ & $0.267 \pm 0.032$ 
\enddata
\end{deluxetable*}

To prepare the data for statistical analysis, each lightcurve was detrended by a low-order polynomial so the mean value is zero. Then, the time series is subsequently renormalized by its own standard deviation, $\sigma_{N}$.
Next, the lightcurves extracted for each observing sequence were examined for the presence of non-Gaussian intensity enhancements such as macroscopic flare signatures following the methodology described by \citetalias{Dillon:2020}. Emission signatures exceeding 3$\sigma_{N}$ above the mean value, lasting continually for a minimum of 1~minute (5~datapoints), were identified in each lightcurve. Based on a normal distribution, the probability of these event presenting through Gaussian-Poisson noise is {$\lesssim2\times10^{-13}$}, hence allowing for robust detection of macroscopic flaring activity. Every star, apart from the M2V candidate NGTS~J062005.7-372555, demonstrated macroscopic flare signatures, resulting in the removal of a further $\sim0.2-2.5$\% of the remaining M-type time series. The degree of macroscopic flare emission varied with the spectral sub-type, with M4V stars exhibiting approximately five times more detected flares than the M0V stellar types, consistent with previous studies \citep{Hawley:2014, Yang:2017}.
Once the larger-scale flare signatures had been identified, they were subsequently removed from the time series using an interval of $\pm$5~minutes ($\pm$25~datapoints) from the first and last detection above the 3$\sigma_{N}$ threshold (see Figure~\ref{Lightcurves}). Removing these signatures allows for the assumption of normality in the intensity distribution, i.e. shot and readout noise combined with ambient stellar intensity \citep{Terrell:1977, Delouille:2008}. This has been shown as valid in \citet{Terzo:2011} and \citet{Jess:2014} alongside \citetalias{jess:2019} and \citetalias{Dillon:2020}, and allows for sensitivity in the detection of any residual intensity deviations within the expected intensity ranges of a normal distribution. The number of macroscopic flares removed were used to calculate approximate flare rates for the M stars, which are displayed in Table~{\ref{tab:StellarParam}}. To ensure consistency with previous stellar nanoflare investigations, the filtering steps employed were identical to those used in \citetalias{Dillon:2020}, with the filtered lightcurves subsequently cropped to 97{\,}060 datapoints each to match the number statistics from the previous study. This allows a direct comparison to be made with the work of \citetalias{Dillon:2020}, since the previously published nanoflare simulations can be re-used due to identical number statistics, filtering techniques, desired $\alpha$ (power-law index) and $\tau$ ($e$-folding time) ranges, in addition to specific NGTS-modeled noise characteristics. 

\section{Analysis and Discussion}
\label{sec:analysis}

To investigate the possible changing nanoflare properties with spectral type, we utilized the statistical and Fourier analysis techniques outlined in Section~\ref{sec:previousanalysis}. As outlined, nanoflares give rise to 2 distinct statistical signatures, which can be used to diagnose stellar nanoflare activity. We present two example histograms of intensity fluctuations in Figure~{\ref{Histogram}} for stars NGTS~J052346.3-361114 (M0V spectral type; top panel) and NGTS~J050423.8-373021 (M4V spectral type; lower panel). From Figure~{\ref{Histogram}} it is clear that opposite ends of the included spectral types, which lie on either side of the predicted fully-convective boundary, demonstrate distinctly different statistical signatures. The M0V star exhibits weak nanoflare signatures, with a marginal negative median offset and no elevated intensity fluctuations at $\sim 2~\sigma_{N}$. On the contrary, the M4V star has a clear excess of $\sim 2~\sigma_{N}$ intensity fluctuations in addition to a prominent negative median offset. The signatures of the M4V star shown in Figure~{\ref{Histogram}} are consistent with previous positive stellar nanoflare identifications in \citetalias{Dillon:2020}. The distinct increase of visible nanoflare signatures within the expected regime of full convection indicates that the enhanced nanoflare rates are related to the underlying convective nature of the star.

\begin{figure*}
   \centering
   \includegraphics[ width=0.8\textwidth]{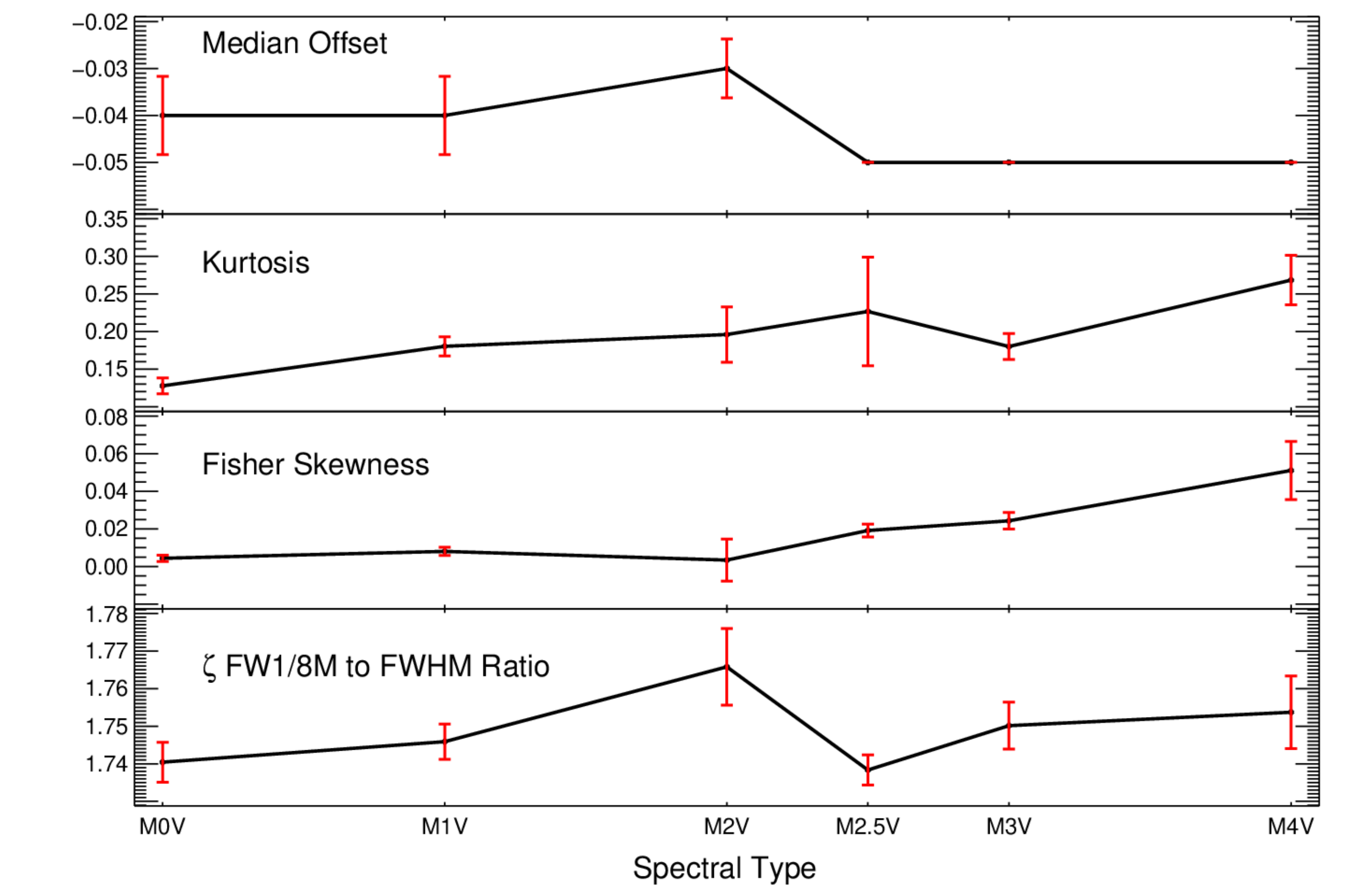}
      \caption{The bootstrap-averaged statistical properties of the intensity fluctuation histograms for each stellar classification. Beyond the convective boundary, at approximately M2.5V and later, sub-types begin to exhibit statistical signatures that are consistent with the presence of nanoflare activity, including larger median offsets (top panel), increasing levels of kurtosis (second panel from top), and higher Fisher skewness values (second panel from bottom). The $\zeta$ (FW$\frac{1}{8}$M-to-FWHM ratio) values do not vary significantly as a function of stellar classification. However, this is likely due to the interplay between the power-law index of the nanoflares and the duration of the $e$-folding timescales, which are able to counteract the statistical effects of one another. }
         \label{AVG_Statistical}
   \end{figure*}

These examples illustrated in Figure~{\ref{Histogram}} clearly identify the vastly different nanoflare signatures present at either end of the investigated range of spectral sub-types. To better examine the change in nanoflare activity across the given spectral range (M0V -- M4V), the derived properties were averaged according to their specific spectral type following the bootstrap method documented by \citet{Efron:1979}. Straightforward averaging of features that are dependent on the underlying stellar plasma conditions from multiple stars is challenging due to the uncertain behavior of the standard errors of the given parameters. Hence, bootstrapping techniques are used extensively throughout the physical sciences to better calculate confidence intervals for data following non-standard or unknown distributions \citep{Simpson:1986,Desmars:2009, Yao:2017}. 

Figure~{\ref{AVG_Statistical}} shows the change in the median offset, kurtosis, Fisher skewness, and $\zeta$ values, respectively, as a function of spectral sub-type. The results are also tabulated in Table ~{\ref{tab:Avg_stats}}. From Figure~{\ref{AVG_Statistical}}, we find a distinct change in the nanoflare statistical signatures as a function of spectral sub-type, suggesting the convective boundary may play an important role in the generation of efficient nanoflare conditions. We find that M2.5V (and beyond) stars exhibit distinct nanoflare statistical signatures that are consistent with those put forward by \citetalias{Dillon:2020}. Specifically, the average median offset for the pre-M2.5V stars exhibits a large spread around a weakly offset value (upper panel of Figure~{\ref{AVG_Statistical}}), while the post-M2.5V stars demonstrate a larger consistent offset magnitude (with less uncertainty) of approximately $-0.05 \sigma_{N}$. 

The Fisher skewness value is effectively zero for pre-M2.5V stars (second panel from bottom in Figure~{\ref{AVG_Statistical}}), suggesting no, or very weak, nanoflare activity. From M2.5V onward, there is a clear increasing trend in the Fisher skewness value of the fluctuation distribution, with the M4V sub-type displaying a Fisher skewness equal to $0.051 \pm 0.014$, providing strong evidence for the presence of nanoflares.

\begin{figure*}
   \centering
   \includegraphics[width=0.8\textwidth]{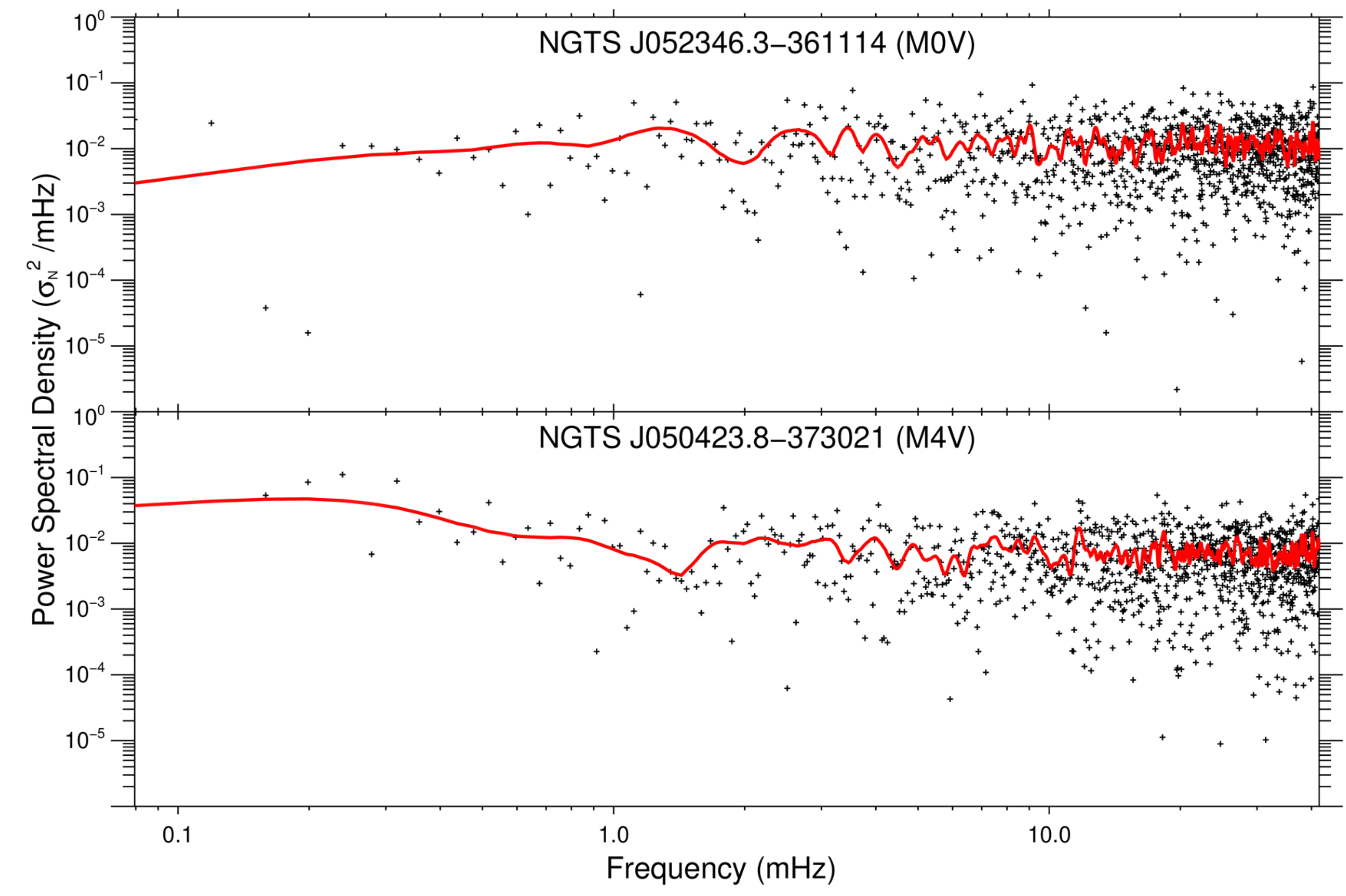}
      \caption{The Fourier power spectral densities (PSDs) for example M0V (upper panel) and M4V (lower panel) stellar sources, displayed in normalized units of $\sigma_{N}^{2}/\text{mHz}$. The crosses in each panel depict the individual power values as a function of frequency, while the solid red line reveals a trendline calculated over $\pm6$ frequency elements ($\pm0.478$~mHz). It can be seen that the PSD for the M0V star is relatively flat, with small-amplitude power enhancements in the range $3-10$~mHz, which is consistent with typical $p$-mode oscillations. On the contrary, the PSD for the M4V star exhibits a clear enhancement of spectral energy at lower frequencies, resulting in a spectral slope of $\beta = -0.57 \pm 0.05$ that begins at $0.32 \pm 0.04$~mHz, followed by numerous power peaks in the range of $1-10$~mHz, which is consistent with the presence of both nanoflare activity and $p$-mode oscillations.}
         \label{PSD_NGTS}
\end{figure*}

In the additional distribution diagnostics, the relationship is less clear. Regarding the kurtosis (second panel from top in Figure~{\ref{AVG_Statistical}}) there appears to be a trend in that statistical kurtosis increases across the full sample, between M0V and M4V. However, the exact nature of this relationship is obscured by the large uncertainty in the kurtosis for spectral types around the dynamo mode transition. In particular, the M3V sub-type appears to exhibit a decrease in kurtosis, though this may be due to the large uncertainty in M2V \& M2.5V producing abnormally large values.

There is no clear trend visible in the corresponding $\zeta$ values (lower panel of Figure~{\ref{AVG_Statistical}}) as a function of spectral sub-type. 
It must be remembered that the $\zeta$ value is a measure of the deviation away from a standard Gaussian distribution, which has a value of $\zeta = 1.73$. As discussed in \citetalias{jess:2019}, increased nanoflare decay timescales (i.e., larger $\tau$ values) result in broader tails of the intensity fluctuation distributions, hence giving rise to $\zeta > 1.73$. On the contrary, large power-law indices help reduce the widths of the tails in the intensity fluctuation distributions due to the superposition of positive intensity fluctuations (e.g., new nanoflares) superimposed on top of decaying (i.e., negative) intensity fluctuations, which result in $\zeta < 1.73$. As such, the interplay between the power-law index and the nanoflare $e$-folding time produces the specific value of $\zeta$ measured. As such, the relatively consistent values of $\zeta$ found across the spectral range M0V -- M4V may result from an increased nanoflare rate expected for M4V stars being negated by an increase in the associated decay timescales of the resulting nanoflares, i.e., a larger $\alpha$ term being coupled with longer $\tau$ values.

As described in Section~\ref{sec:previousanalysis} (and in detail in \citetalias{jess:2019}), the products of the four distribution diagnostics can be used to derive the power-law index, $\alpha$, and the nanoflare decay timescale, $\tau$, for each observation. Through the calculation of median offset, Fisher skewness, kurtosis, and $\zeta$ for each of the $6100$ synthesized lightcurves, these can be compared to the values seen derived in Table~\ref{tab:Avg_stats}. This was achieved by considering each diagnostic individually. The values of the statistical diagnostic within a range of $\pm 1 \sigma_{N}$ are directly compared to the corresponding simulated signatures, to determine which $\alpha$ and $\tau$ values match. The values of these nanoflare paramaters which equal all four of the observed diagnostics are the derived $\alpha$ and $\tau$ values. For pre-M2.5V stars, it was not possible to establish values for the power-law index and $e$-folding time that were self-consistent with the Monte Carlo models provided by \citetalias{Dillon:2020}. For example, it was possible to find self-similarity between the observational and model power-law indices, but this resulted in decay timescales that were incompatible and inconsistent. As a result, we are unable to define nanoflare characteristics for pre-M2.5V stars, suggesting that nanoflare activity may be very weak (or not present) on these specific stellar sub-types.

\begin{deluxetable}{lccc}[!t]
\label{tab:Stat_Alpha}
\tablecaption{Nanoflare parameters per spectral type, derived from statistical properties of Monte-Carlo modeled nanoflare timeseries. The approximately symmetrical distribution of statistical properties leads to an ambiguity in the derived power-law indices, hence $\alpha_1$ and $\alpha_2$. }
\tablewidth{0pt}
\tablehead{
\colhead{Spectral type} &   \colhead{ $\alpha_1 $ }&   \colhead{ $\alpha_2 $ }  & \colhead{ $\tau (s)$} 
}
\startdata
M2.5V&  $2.25 \pm 0.25$ & $3.00\pm 0.25$ & $200 \pm 100$\\
M3V &   $2.25 \pm 0.20$ & $3.00\pm 0.20$ & $200 \pm 100$\\
M4V &   $2.30 \pm 0.20$ & $3.10\pm 0.20$ & $450 \pm 50$ \\
\enddata

\end{deluxetable}

\begin{figure*}
   \centering
   \includegraphics[ width=0.8\textwidth]{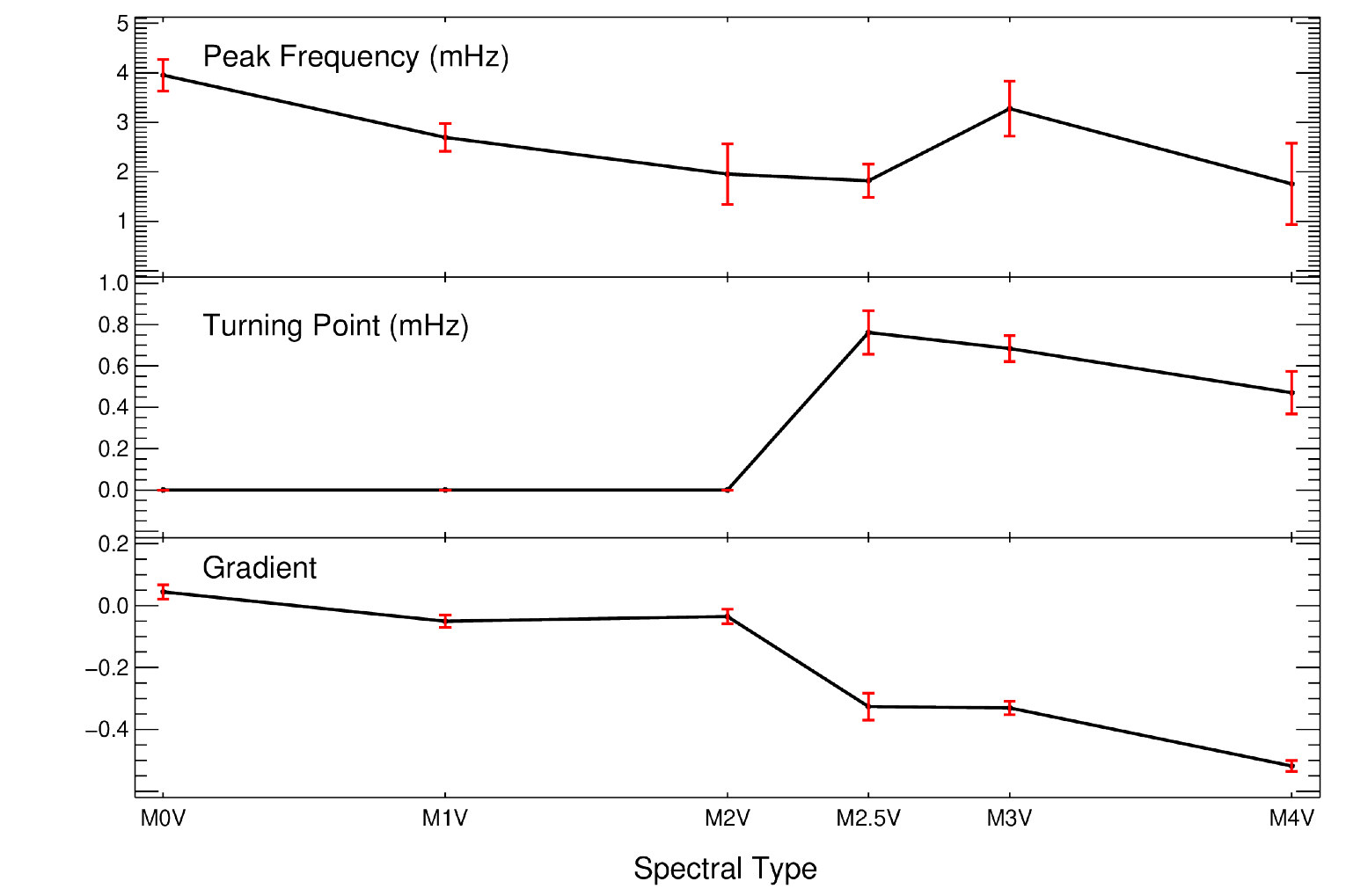}
      \caption{The bootstrap-averaged properties of the Fourier power spectral densities (PSDs) across each spectral type. The upper panel displays the peak frequency values (in mHz), which are found to reside within the range of approximately $1-4$~mHz, which is consistent with both nanoflare activity and $p$-mode oscillations, and therefore cannot be used as an indicator of nanoflare activity by itself. The middle and lower panels display the turning point frequencies (in mHz) and subsequent spectral slopes, respectively, as a function of stellar classification. When compared to the Monte Carlo nanoflare simulation outputs depicted in Figure~{\ref{HeatMapReGen}}, the distinct jump in turning point frequency and spectral gradient at the convective boundary (M2.5V) provides clear evidence of prominent nanoflare activity in M2.5V -- M4V stellar sources. }  
         \label{AVG_Fourier}
   \end{figure*}

The statistical parameters for the M2.5V, M3V and M4V stars, which are believed to be beyond the convective boundary and therefore best described as `fully convective', exhibit values consistent with the power-law indices of $\alpha=2.25 \pm 0.25$ or $\alpha=3.00 \pm 0.25$, $\alpha=2.25 \pm0.20$ or $\alpha=3.00 \pm 0.20$,  and $\alpha=2.30 \pm0.20$ or $\alpha=3.10 \pm 0.20$, alongside the $e$-folding timescales of $\tau=200 \pm 100$~s, $\tau=200 \pm100$~s, and $\tau=450 \pm50$~s, respectively (see Table~{\ref{tab:Stat_Alpha}}). As highlighted by \citetalias{Dillon:2020}, the approximate symmetry of the statistical distributions about their peak values leads to ambiguity in the derived power-law indices (see, e.g., the bands of similar values shown in each panel in Figure~{5} of  \citetalias{Dillon:2020} ). As a result, it is possible to map each sub-type onto two distinct solutions for the power-law index. Irrespective of this ambiguity, both sets of possible nanoflare conditions are highly active (i.e., $\alpha>2$), in stark contrast to the effectively zero statistical nanoflare signals observed in the pre-M2.5V spectral sub-types. The larger uncertainty in the M2.5V power-law indices are due to the larger uncertainty associated with the kurtosis value for these spectral sub-types. M2.5V stars are at the boundary of predicted full convection, so a larger spread in their nanoflare properties would be expected if full convection is the cause of the spectral `break' in associated power-law indices.

\begin{deluxetable*}{ccccc}[]
\label{tab:Avg_Fourier}

\tablecaption{Average characteristics of the Fourier PSD properties by each spectral type. }

\tablewidth{0pt}
\tablehead{
\colhead{~Spectral type~} &   \colhead{ Gradient }  & \colhead{Turning point (mHz)}& \colhead{ Peak frequency (mHz) } 
}
\startdata
M0V &   $ +0.044 \pm 0.023$ & $0.000 \pm 0.000$ & $3.952 \pm 0.320$\\
M1V &   $-0.051 \pm 0.019$ & $0.000 \pm 0.000$ & $2.695 \pm 0.278$\\
M2V &   $-0.035 \pm 0.022$ & $0.000 \pm 0.000$ & $1.956 \pm 0.611$\\
M2.5V&  $-0.326 \pm 0.044$ & $0.762 \pm 0.105$ & $1.821 \pm 0.338$\\
M3V &   $-0.330 \pm 0.022$ & $0.684 \pm 0.063$ & $3.276 \pm 0.557$\\
M4V &   $-0.518 \pm 0.018$ & $0.467 \pm 0.103$ & $1.757 \pm 0.822$\\
\enddata

\end{deluxetable*}

Interestingly, the M4V stars display evidence for longer $e$-folding timescales when compared to their M2.5V and M3V counterparts. This may imply that the power-law index is marginally greater than for the earlier spectral classes. As previously discussed, constant $\zeta$ values are seen throughout the spectral sub-type range, and are thought to be due to the statistical effects of larger power-law indices being negated by the slower decay timescales associated with those stars (see \citetalias{jess:2019} for a more thorough discussion of this interplay). The specific values for the $e$-folding timescales for the M2.5V and M3V stars of $\tau = 200\pm100$~s, are consistent with the previous work of \citetalias{Dillon:2020}, who studied similar stellar types. 

Overall, the changes in the statistical parameters indicate that post-dynamo mode transition M-dwarf stars (i.e., M2.5V and later and fully convective) exhibit greatly enhanced stellar nanoflare activity when compared to the partially convective pre-dynamo mode transition M-dwarfs that show little-to-no evidence for nanoflare activity.

As highlighted in \citetalias{Dillon:2020}, the examination of Fourier signatures, which are derived directly from the stellar lightcurves, can help disambiguate any derived nanoflare characteristics and further substantiate the evidence for specific activity levels. Following the methods documented by \citet{Welch:1961} and \citet{Vaughan:2012}, power spectral densities (PSDs) were derived from the stellar time series. The longest continuous time series (i.e., the longest uninterrupted series of frames) common to all stars was $2095$~datapoints, slightly shorter than the $2316$ consecutive frames employed by \citetalias{Dillon:2020}. This resulted in the frequency resolution being slightly reduced from $\Delta{f} = 0.0356$~mHz to $\Delta{f} = 0.0398$~mHz in the present study. In order to readily compare the observational PSDs to those calculated from the Monte Carlo nanoflare models of \citetalias{Dillon:2020}, the Fourier signatures needed to be re-calculated adhering to the new frequency resolution. Hence, utilizing the new frequency resolution, we re-computed the PSDs and corresponding `heat map' of the simulated Fourier properties \citep[c.f., Figure~7 of][]{Dillon:2020} as a function of both the nanoflare power-law and $e$-folding time. The recalculated heat map is displayed in Figure~{\ref{HeatMapReGen}}. Due to the change in frequency resolution being a relatively small value ($0.0042$~mHz), no noticeable deviations from Figure~{\ref{HeatMapReGen}} and the original distribution \citep[Figure~7 of][]{Dillon:2020} can be seen, with the trends identical in both studies. This is likely a result of the Fourier nanoflare trends being relatively broad in frequency with peak values sufficiently away from the lowest and highest (i.e., Nyquist) frequencies of the time series, and therefore are not significantly affected by very small changes in the underlying frequency resolution.

As with the statistical signatures shown in Figure~{\ref{AVG_Statistical}}, there are dramatic differences in the Fourier properties between M0V and M4V stars. As seen in Figure~{\ref{PSD_NGTS}}, the M0V has an effectively flat power spectrum \citep[suggesting no nanoflare signal is present;][]{Dillon:2020}, which is contrasted by the M4V star that demonstrates a spectral slope of ${\beta = -0.57 \pm 0.05}$ between the frequencies ${\sim0.3-6.0}$~mHz. In Figure~{\ref{PSD_NGTS}}, the black crosses represent the individual frequency-dependent power measurements, while the solid red line depicts a trendline established over $\pm6$ frequency elements ($\pm0.478$~mHz). In the lower panel of Figure~{\ref{PSD_NGTS}}, a PSD slope is consistent with enhanced rates of stellar nanoflare activity, which begins at the `turning point' of $0.32 \pm 0.04$~mHz. As defined by  \citetalias{Dillon:2020}, the turning point is defined as the initial peak before the gradual reduction in Fourier power with increasing frequency. It must be noted that both PSD plots shown in Figure~{\ref{PSD_NGTS}} (i.e., for M0V and M4V spectral types) exhibit numerous power peaks in the range of $1-10$~mHz, consistent with both stellar nanoflare signatures \citep{Dillon:2020} and the presence of $p$-mode oscillations generated in the convective layers of M-dwarf stellar sources \citep[M-dwarf stars are believed to exhibit solar-like oscillations, hence $p$-modes synonymous with the typical solar frequency range;][]{Rodriquez:2014, Rodriguez:2016}.  As the entire range of spectral types included in this study (M0V -- M4V) are expected to exhibit $p$-mode oscillations, the peak frequencies within this interval are not conclusive evidence alone of nanoflare activity. 

The averaged (following bootstrap procedures) Fourier properties per spectral type are shown in Figure~{\ref{AVG_Fourier}}, and tabulated in Table~{\ref{tab:Avg_Fourier}}. As with the averaged statistical signatures shown in Figure~{\ref{AVG_Statistical}}, there is a marked change in Fourier features consistent with nanoflare activity for spectral classifications M2.5V and later. Evidence for this is shown in the averaged PSD spectral gradient (lower panel of Figure~{\ref{AVG_Fourier}}), where pre-M2.5V stars have relatively flat spectral slopes ($\beta \sim 0$), yet stellar sources past the convective boundary at M2.5V and later demonstrate increased magnitude spectral slopes in the range of $-0.6 \leq \beta \leq -0.3$. Note that the peak frequency values (upper panel of Figure~{\ref{AVG_Fourier}}) are relatively consistent across all M-dwarf stellar sources, approximately in the range of $2-4$~mHz. As discussed above, this alone does not constitute evidence of nanoflare activity since all of these sources are expected to demonstrate $p$-mode oscillations spanning that particular frequency interval \citep{Guenther2008,Rodriquez:2014}.

The corresponding `turning point', where the spectral slopes are observed to begin, is, of course, equal to zero for the pre-M2.5V stars since they do not exhibit any associated spectral slopes (middle panel of Figure~{\ref{AVG_Fourier}}). However, for spectral classifications beyond M2.5V, where the stars are believed to be fully convective, a relatively constant value (when errors are included) in the range of ${0.3 \leq f \leq 0.9}$~mHz is found, which is consistent with the previous work of \citetalias{Dillon:2020}. In simulated nanoflare lightcurves documented by \citetalias{Dillon:2020}, an increased flare decay rate (i.e., a longer $\tau$ value) gave rise to a decreased frequency of the Fourier turning point. Examination of the middle panel of Figure~{\ref{AVG_Fourier}} shows that while the turning point frequencies are distinctly different from the pre-M2.5V stars, there does seem to be tentative evidence that the average turning point frequency decreases across the M2.5V, M3V, and M4V spectral types. This is further evidenced in Table~{\ref{tab:Avg_Fourier}}, where the turning points of the M2.5V, M3V, and M4V stars are computed as $0.762\pm0.105$~mHz, $0.684\pm0.063$~mHz, and $0.467\pm0.103$~mHz, respectively. The evidence suggests that the $e$-folding timescales associated with the M4V stars are longer than their M2.5V counterparts, which is consistent with the intensity fluctuation statistical signatures discussed above.

\begin{deluxetable}{lcc}[]
\label{tab:Fourier_Alpha}
\tablecaption{Nanoflare parameters per spectral type, derived from Fourier properties of Monte-Carlo modeled nanoflare timeseries. There is no ambiguity in the derived power-law indices. }
\tablewidth{0pt}
\tablehead{
\colhead{Spectral type} &   \colhead{ $\alpha $ }  & \colhead{ $\tau (s)$} 
}
\startdata
M2.5V  & $3.00\pm 0.15$ & $200 \pm 100$\\
M3V    & $3.00\pm 0.15$ & $250 \pm 100$\\
M4V    & $3.10\pm 0.15$ & $450 \pm 50$\\ 
\enddata

\end{deluxetable}

Comparing the derived Fourier properties to the heat maps shown in Figure~{\ref{HeatMapReGen}}, it is possible to estimate the power-law indices and decay timescales for each of the M2.5V, M3V, and M4V stellar types that shown clear evidence for nanoflare activity. We find power-law indices of $\alpha = 3.00 \pm  0.15$, $\alpha = 3.00 \pm  0.15$, and $\alpha = 3.10 \pm  0.15$, alongside nanoflare $e$-folding timescales of $\tau = 200 \pm 100$~s, $\tau = 250 \pm 100$~s, and $\tau = 450 \pm 50$~s, for the M2.5V, M3V, and M4V spectral types, respectively (see Table~{\ref{tab:Fourier_Alpha}}). Importantly, these values are consistent with the statistical analyses, with the Fourier techniques providing additional benchmarks to validate the nanoflare properties extracted from the observational time series and resolve the ambiguity in power-law index arising from the statistical analysis. In contrast to the statistical mapping, the derived Fourier parameters of the M3V stars are consistent with a marginal $e$-folding time enhancement compared to the M2.5V classifications. This is likely related to the same physical processes that caused enhanced $e$-folding timescales in the M4V star. However, this is difficult to ascertain due to the relatively large errors in determining the plasma decay rate over the entire stellar surface.

Combining the Fourier and statistical analyses (see Table~{\ref{tab:Combined_Alpha}}), we find that the fully convective M2.5V and M3V sub-types exhibit nanoflare power-law indices of $\alpha = 3.00 \pm  0.20$ and $\alpha = 3.00 \pm  0.18$, respectively. The M2.5V sub-types are consistent with a decay timescale of $\tau = 200 \pm  100$~s, whereas the M3V stars display tentative evidence for a slightly enhanced $e$-folding timescale of $\tau = 225 \pm  100$~s. These $e$-folding timescales and power-law indices are values consistent with similar M-dwarf spectral types studied by \citetalias{Dillon:2020}, whereas M4V stars exhibit elevated power-law indices of $\alpha = 3.10 \pm  0.18$, with an increased decay timescale of $\tau = 450 \pm  50$~s. With these properties confirmed, the behavior of these flares in comparison to M-dwarf flare samples as-a-whole can be inferred. It has been established that a general relationship between the flare duration, $t$, and energy holds for observable flare populations, namely that $t \propto E^{x}$, where $x\approx0.33$ for solar and G-type stellar flares \citep{2002A&A...382.1070V,2015EP&S...67...59M}, whereas it drops to $x\approx0.2$ for solar microflares \citep{2008ApJ...677.1385C}. In \citet{2015ApJ...814...35C}, a directly comparable relationship between the $e$-folding time (in minutes) and the flare energy was defined from a sample of 420 energetic M-dwarf flares ($E \simeq 10^{31} - 10^{34}$~erg). The log-log fit of the data was established at a high statistical significance as,
\begin{equation}
log~\tau = (0.57 \pm 0.05)~log~E - (15.61 \pm 1.57) \ . 
\label{logfit}
\end{equation}
Taking the peak energies for nanoflares as $E = 10^{25}$~erg, Equation~\ref{logfit} produces $e$-folding times of $1 \leq \tau \leq 258$~s. From inspection, the derived $e$-folding times of the M2.5V and M3V populations are consistent with the upper boundary of predicted values, whereas the M4V values lie outwith the derived relationship. This is not necessarily unexpected, since a complementary study by \citet{2019ApJ...881....9H} found a broken power-law index relationship between the $e$-folding time and flare energy, where at $E \leq 10^{33}$~erg, $\tau$ remained approximately constant instead of following the trend associated with Equation~\ref{logfit}. The authors attributed this to the limitations of flare characterizations around the detection limit, but the flares in this study suggest that the effect may be physical. Equation~\ref{logfit} was derived from flare energies orders of magnitude above those under consideration in this study, thus the disparity between the predicted $e$-folding time and that seen in M4V stars may be indicative of a transition from large-scale Petschek reconnection to the Sweet-Parker process. As was proposed by \citet{Tsuneta:2004}, small-scale pico/nano-scale flares occur more favorably via Sweet-Parker than Petschek reconnection. As \citetalias{Dillon:2020} suggest, this would explain a discontinuity in the power-law relationship between nanoflares and their larger scale counterparts, which remain driven by Petschek-like reconnection \citep{Loureiro:2016}. The Sweet-Parker reconnection process is inversely proportional to the square root of the plasma Lundquist number, which is itself inversely proportional to the plasma resistivity. As such, Sweet-Parker reconnection is more favorable in poorly conducting plasmas. The increased decay timescale of  $\tau = 450 \pm  50$~s, alongside the associated increased power-law index of $\alpha = 3.10 \pm  0.18$, found for the M4V sub-type may be related to increased plasma resistivity, which matches expectations for mid-to-late M-dwarfs \citep{Mohanty:2002}. 
Caution is required however ; these increased $\alpha$ values are within 1 $\sigma_{N}$, and the $\tau$ values within 3 $\sigma_{N}$ of the uncertainties of the less enhanced M2.5 and M3V stars, so this trend cannot yet be considered statistically significant. Future investigation of M5V and later sub-types is required to determine if there is a statistically significant trend exceeding 3 $\sigma_{N}$ confidence in the observed properties. This could be complimented by multi-color observations that would allow for lower uncertainty in the $\tau$ value at each color band due to the reliance on underlying plasma properties, which are naturally more separated across color bands due to their associated temperature sensitivities. 

In contrast to the fully convective sub-types, pre-dynamo mode transition M0V -- M2V stars exhibited weak (if any) nanoflare signals, suggesting that fully convective stellar atmospheres lead to a large enhancement of nanoflare activity.

While the observed trend of fully convective stars exhibiting enhanced nanoflare activity is clear, the exact mechanism leading to this is still a matter of debate. While the Sweet-Parker hypothesis is plausible, there is also a potential issue. If enhanced nanoflare activity occurs in the corona, it would lead to enhanced heating of that plasma. Consequently, this would lower the resistivity and hence lower the rate of Sweet-Parker reconnection. This `feedback loop' behavior may reach some natural and stable equilibrium, but it may be necessary to incorporate additional theory to ensure the stability of this mechanism. Referring to the original nanoflare mechanism theorized by \citet{Parker:1988} may provide this. In that paper, Parker suggested random convective motion in the photosphere causes `shuffling' and subsequent deformation and braiding of the photospheric footpoints of the coronal magnetic fields and consequently the generation of free energy. The coupling of the magnetic field lines between the photosphere and corona provides the framework to allow this free energy to flow into the corona. This energy is then dissipated in coronal current sheets, leading to small-scale reconnection. As such, enhanced heating leading to decreased resistivity would improve the magnetic coupling between these footpoints and the corona, consequently enhancing the flow of free energy available for nanoflare activity. One can imagine a combination of these scenarios, wherein the sympathetic transfer of hot plasma and free energy through these coupled fields regulates the resistivity and drives a stable rate of Sweet-Parker reconnection. 

To uncover the source of this enhanced activity, it’s vital to obtain two sets of observations; multi-band photometry, and observations of later MV star types. The multi-color observation of these stars will allow us to make a limited analysis of the change in nanoflaring properties across different wavelengths and consequently the contribution at different atmospheric heights. Comparing relative photospheric and coronal signatures could diagnose the underlying mechanism powering this enhanced nanoflare activity. The multi-color analysis should also provide a lower uncertainty in the $\tau$ values. Secondly, sourcing  M5V and beyond stars would allow the continuation of the trend in flare decay rate (if any) to be investigated. If later MV stars continue to exhibit enhanced activity it would support the Sweet-Parker reconnection theory, as it would suggest the enhanced resistivity is key. Ultimately, observations of later MV stars, and across multiple photometry bands will need to be coupled with detailed physical modeling to try and uncover what changes in these stars are driving their nanoflare behavior.

Regardless of the specific physical mechanism causing this enhancement across the convective boundary, it is there. The observational evidence points to nanoflare contributions increasing significantly in the fully convective M2.5V and later stars. This novel result is independent of the modeled nanoflare lightcurves, which serve only to diagnose the parameters of the nanoflare signatures within observed lightcurves. It is also independent of the range of stellar luminosities present in the sample. It is established that greater macroscopic flare rates in later-type stars, as seen in Table~\ref{tab:StellarParam}, are influenced by the reduced luminosity threshold of these stars. It is therefore of interest to investigate whether the reduced flare detection threshold may influence the nanoflare study presented here. Average luminosities for the M0 and M4 stars, representing the largest range of spectral classes under consideration, are $L_{M0} = 0.068 L_{\odot}$ and $L_{M4} = 0.014 L_{\odot}$. Therefore, the energy rates associated with these luminosities can be estimated as $E_{M0} = 2.6\times10^{32}$~erg~s$^{-1}$ and $E_{M4} = 5.5\times10^{31}$~erg~s$^{-1}$, respectively. This order-of-magnitude drop in the energy rate associated with the fundamental stellar brightness agrees with previously detected flare energies \citep[e.g.,][]{Martinez2020}. In the present study, a $1\sigma_{N}$ deviation is modeled in the simulation as $5\times10^{24}$~erg \citep{jess:2019, Dillon:2020}, and by projecting an order-of-magnitude energy threshold decrease between M0 and M4 classifications to nanoflare conditions, $1\sigma_{N}$ deviations in M4V stars equates to flares with energies $\sim10^{23}$~erg, still well within the typical nanoflare regime. This differential in flare energy sampling is also not sufficient to explain the lack of nanoflare enhancement in early-type MV stars. The flare frequency, $dN/dE$ from Equation~\ref{eqn:powerlaw}, associated with the M2.5V nanoflares calculated from the $\alpha$ and $\tau$ values seen in Table~\ref{tab:Combined_Alpha} reveals a two order-of-magnitude increase in nanoflare frequency between M0 and M4 ($10^{-46} - 10^{-44}$~erg$^{-1}$ cm$^{-2}$ s$^{-1}$), consistent with previous solar studies \citep{2022A&A...661A.149P}. Therefore, in a scenario where early-type M-dwarfs had the same flaring profile as their later-type counterparts, there would still be an ample frequency of 1$\sigma_{N}$ signatures present in their histograms and Fourier spectra. Given that there is no such behavior seen, we posit this as evidence that the nanoflare frequency spectrum detected in this work exists only in the fully-convective sample. Finding the source of this convective divide should be a key focus of future studies.

\begin{deluxetable}{lcc}[]
\label{tab:Combined_Alpha}
\tablecaption{Nanoflare parameters per spectral type, derived from combined statistical and Fourier properties of Monte-Carlo modeled nanoflare timeseries. }
\tablewidth{0pt}
\tablehead{
\colhead{Spectral type} &   \colhead{ $\alpha $ }  & \colhead{ $\tau (s)$} 
}
\startdata
M2.5V  & $3.00\pm 0.20$ & $200 \pm 100$   \\
M3V    & $3.00\pm 0.18$ & $225 \pm 100$  \\
M4V    & $3.10\pm 0.18$ & $450 \pm 50$  \\
\enddata

\end{deluxetable}

The enhanced small-scale flare rates in fully convective stars holds profound implications for the energy budgets of those stellar sources. The energy output of rapid and continuous nanoflares may be a major component of the overall stellar energy budget, yet are hidden within the noise envelope of the observations and can only be extracted through use of large-scale statistical and Fourier analyses. The question of whether the enhanced flaring visible in post-dynamo mode transition M2.5V -- M4V stars is due to the helical dynamo or altered plasma Lundquist conditions in these stars is an avenue to explore in future work. Furthermore, our work reveals tentative evidence that M4V stars are linked to nanoflare events that have inherently longer decay timescales (i.e., larger $\tau$ values) as well as larger power-law indices. Importantly, mid M-dwarf sub-types should have decreased optical depths, alongside increased plasma resistivities, a trend which continues to late M9 sub-types \citep{Mohanty:2002}. If the nanoflare $e$-folding times continue to increase with increasing M-dwarf sub-type, it would support the scenario of increased plasma resistivity leading to increased small-scale flaring via Sweet-Parker reconnection. This would appear to support the findings of \citet{Wright:2016,Wright:2018}, that solar and stellar dynamos operate independent of a tachocline. As a result, it is of paramount importance to source sufficient late M-type stellar time series for follow-up analyses. 

\section{Conclusions}

Evidence for stellar nanoflares has been observed on a further 15 post-dynamo mode transition (M2.5V, M3V, and M4V classification) stars, with nanoflare power-law indices and $e$-folding times consistent with the enhanced rates of nanoflare activity put forward by \citetalias{Dillon:2020}. The marked increase in nanoflare activity is coincident with M2.5V and later sub-types, suggesting that the change from partial to fully convective atmospheres may be responsible. The post-dynamo mode transition stars exhibit nanoflare rates that are enhanced from those seen at larger energies in other stars and the Sun, with power-law indices found to be in the region of $\alpha = 3.00 \pm 0.20$ for M2.5V and M3V sub-types, with slightly larger values of $\alpha = 3.10 \pm  0.18$ for M4V sub-types. Given the relation between power-law index and low-energy flare frequency, it is clear that the atmospheres of late MV stars have an energy budget dominated by small-scale flaring. Whereas observational evidence of nanoflares being the dominant heating mechanism in the solar atmosphere remains elusive, their energy output in fully-convective stars may well be sufficient to produce bulk heating. The decay timescales for M2.5V and M3V stars were found to be on the order of $\tau = 200 \pm  100$~s, while evidence was presented for increased plasma $e$-folding times of ${\tau = 450 \pm  50}$~s in the M4V stars, suggesting the presence of Sweet-Parker reconnection processes. It must be noted these enhanced values for the M4V star remain within  $1-3\sigma_{N}$ of the M2.5 and M3V stars, so we cannot yet consider these to be fully distinct.

On the contrary, pre-dynamo mode transition M-dwarf (M0V, M1V, and M2V classification) stars exhibit marginal statistical or Fourier-based nanoflare signals, indicating that the large power-law index for MV stars reported in \citetalias{Dillon:2020} is not uniform across all spectral types. Instead, it is implied that a fully-convective interior is necessary to exhibit the $\alpha \geq 3$ that distinguish them from other stellar candidates. Additionally, the underlying reason why fully convective atmospheres lead to enhanced nanoflare activity should be explored, i.e., is this due to an altered dynamo, or due to other plasma changes such as modification of the corresponding Lundquist number? One avenue of exploration would be examining M5V (and later) stellar types, to investigate if there is a continuing and more statistically significant trend in the flare decay rate and associated power-law index, which could be linked to increasing plasma resistivity, and thus increased Sweet-Parker reconnection rates. It is likely such observations would need to be coupled to detailed theoretical and modeling efforts using well-developed numerical simulations \citep[e.g.,][]{Takahashi:2011, Tenerani:2015, Shi:2018, Papini:2019}. 

Additionally, sampling late-type M dwarfs may reveal the traditional observational signatures of nanoflares. Since the advent of M-dwarf flare studies following the seminal works of \citet{1972Ap&SS..19...75G} and \citet{1976ApJS...30...85L}, it has been clear that the higher luminosity of early-type MV stars can skew flare population studies in favor of less luminous later-type stars due to a lower intensity threshold for flare detection. This has led to subsequent studies constraining the population of stars under consideration, or studying a single star in detail \citep[e.g.,][]{2014ApJ...797..121H, 2014ApJ...797..122D}. This is the case in our present study, where we limited the population of MV stars to those with comparable magnitudes. However, utilizing the reduction in surface temperature in late-type MV stars would allow for the intensity threshold of a macroscopic flare detection to be reduced. A sample of MV stars with varying magnitudes would allow for the $2\sigma_{N}$ nanoflare intensity excursions found in this study to potentially lie above the minimum observable detection threshold in cooler late-type MV stars. This could confirm the temporal morphology and occurrence frequencies of flares at these energies and provide further insight into the validity of previously defined relationships, such as Equation~\ref{logfit}. Additionally, the statistical techniques employed here could provide the first signatures of even smaller flares, such as the proposed pico-flare energy regime \citep{2001ApJ...557..343K, 2003PASJ...55.1025K}.  

It goes without saying that enhanced small-scale reconnection in fully convective stars may mean that nanoflare activity could be a significant component of their overall energy budget. Large-scale multi-year studies of stellar nanoflare rates in fully convective M-dwarfs would further our understanding of nanoflare behavior across different activity cycles, which would further shine light on the ubiquity and role nanoflares play in these dynamic host stars. This can be achieved through further use of large-scale sky surveys (like the NGTS) and space-based observations from the likes of the Transiting Exoplanet Survey Satellite \citep[TESS;][]{Ricker:2014}, alongside targeted campaigns using high-cadence observational platforms, such as HiPERCAM \citep{Dhillon2016}, or multi-band photometry such as the Rapid Eye Mount (REM) telescope \citep{Antonelli:2005} to investigate the nanoflare signature across layers of the Stellar atmosphere.


\acknowledgments
\noindent We thank the anonymous referee for their valuable input in improving this manuscript.
S.D.T.G, D.B.J., and C.J.D. wish to thank Invest NI and Randox Laboratories Ltd. for the award of a Research and Development Grant (059RDEN-1) that allowed the computational techniques employed to be developed.
D.B.J. and S.D.T.G. also acknowledge support from the UK Space Agency for a National Space Technology Programme (NSTP) Technology for Space Science award (SSc~009). 
S.D.T.G., D.B.J., and M.M. would like to thank the UK Science and Technology Facilities Council (STFC) for the consolidated grants ST/T00021X/1 and ST/X000923/1.
D.B.J. would like to thank the STFC for an Ernest Rutherford Fellowship (ST/K004220/1), in addition to a dedicated standard grant (ST/L002744/1) that allowed this project to be started. 
D.B.J. also wishes to thank The Leverhulme Trust for grant RPG-2019-371.  
C.A.W. acknowledges support from STFC consolidated grant ST/X00094X/1.
J.A.G.J. acknowledges support from grant HST-GO-15955.004-A from the Space Telescope Science Institute, which is operated by the Association of Universities for Research in Astronomy, Inc., under NASA contract NAS 5-26555. 
S.L.C would like to thank STFC for an Ernest Rutherford Fellowship (ST/R003726/1).
P.J.W., D.R.A., and R.G.W. acknowledge support from STFC consolidated grants ST/L000733/1 and ST/P000495/1.
This project is based on data collected under the NGTS project at the ESO La Silla Paranal Observatory. 
The NGTS facility is operated by the consortium institutes with support from the UK STFC under projects ST/M001962/1 and ST/S002642/1. 
This research has made use of data obtained from the 4XMM XMM-Newton Serendipitous Source Catalog compiled by the 10 institutes of the XMM-Newton Survey Science Centre selected by ESA.
J.I.V. acknowledges support of CONICYT-PFCHA/Doctorado Nacional-21191829.
Finally, S.D.T.G. and D.B.J. wish to acknowledge scientific discussions with the Waves in the Lower Solar Atmosphere (WaLSA; \href{https://www.WaLSA.team}{https://www.WaLSA.team}) team, which has been supported by the Research Council of Norway (project no. 262622), The Royal Society \citep[award no. Hooke18b/SCTM;][]{2021RSPTA.37900169J}, and the International Space Science Institute (ISSI Team~502). 
%

\facilities{Next Generation Transit Survey (NGTS)}





\clearpage

\appendix

\setcounter{table}{0}
\renewcommand{\thetable}{A\arabic{table}}
\section{Stellar Parameters}
\label{sec:Appendix_A}
Additional stellar parameters, including the RA and Dec for each star are described in Table~{\ref{tab:StellarParam}}. 
\begin{sidewaystable*}[h]
\centering
\resizebox{\linewidth }{!}{%
\begin{tabular}{cccccccccccc}
\hline
\toprule

Sp Type & NGTS ID               & GAIA ID     & TIC ID    & RA         & Dec        & Mass ($M_\odot$)     & Radius ($R_\odot$)   & Luminosity ($L_\odot$)  & Distance (pc) & Approximate Flare Rate per Hour & Magnitude \\ \hline
M0V     & NGTS J233315.1-385757 & 6538313140873424640 & 224245757 & 353.312913 & -38.965817 & 0.487596 & 0.489513 & 0.04772778  & 100.33        & 0.0123635         & 13.10 \\ 
M0V     & NGTS J045221.8-312424 & 4874911889552910000 & 1310695   & 73.090834  & -31.406834 & 0.597557 & 0.611732 & 0.07879962  & 137.421       & 0.0154544         & 13.26 \\ 
M0V     & NGTS J052346.3-361114 & 4822374303400198144 & 167745038 & 80.94287   & -36.187338 & 0.566358 & 0.574022 & 0.07396496  & 123.332       & 0.00618174        & 13.04 \\ 
M0V     & NGTS J061346.1-362248 & 2885025813007881728 & 267248553 & 93.442125  & -36.380098 & 0.585147 & 0.596407 & 0.07490093  & 137.363       & 0.0123635         & 13.24 \\ 
M0V     & NGTS J061054.6-370701 & 2884885281677800448 & 300200809 & 92.72739   & -37.116954 & 0.575812 & 0.585163 & 0.06398756  & 112.406       & 0.0401813         & 12.99 \\ 
M1V     & NGTS J233248.3-382456 & 6538532356004046592 & 224244565 & 353.201262 & -38.415564 & 0.542372 & 0.546805 & 0.05344092  & 90.7535       & 0.0123635         & 12.97 \\ 
M1V     & NGTS J051250.6-361938 & 4821058497219315328 & 14173066  & 78.210928  & -36.327354 & 0.596268 & 0.61012  & 0.06084004  & 201.14        & 0.0123635         & 14.16 \\ 
M1V     & NGTS J052652.1-373123 & 4821222942926810752 & 192785958 & 81.717213  & -37.523125 & 0.567364 & 0.575197 & 0.05580399  &               & 0.00309087        & 13.57 \\ 
M1V     & NGTS J235034.7-373312 & 2310510165491596672 & 183536494 & 357.644544 & -37.553375 & 0.506923 & 0.509045 & 0.04646323  & 83.9291       & 0.00927262        & 12.84 \\ 
M1V     & NGTS J111257.7-331216 & 5403344977522967424 & 23438898  & 168.240279 & -33.20455  & 0.460039 & 0.462724 & 0.03412572  & 73.1095       & 0.0185452         & 12.92 \\ 
M2V     & NGTS J045136.3-321720 & 4874656837214833664 & 1309522   & 72.901424  & -32.288803 & 0.56     & 0.806538 & 0.1124733   & 247.955       & 0.0278178         & 14.32 \\ 
M2V     & NGTS J050254.6-352000 & 4825066629419253632 & 1526841   & 75.72729   & -35.333409 & 0.469279 & 0.471581 & 0.03436859  & 69.7222       & 0.0185452         & 12.72 \\ 
M2V     & NGTS J051926.5-253444 & 2957763042671388416 & 30960826  & 79.860208  & -25.578811 & 0.460925 & 0.463568 & 0.03302738  & 90.6288       & 0.00927262        & 13.49 \\ 
M2V     & NGTS J053614.4-353309 & 4821870486556489216 & 24612475  & 84.059908  & -35.552443 & 0.501857 & 0.503859 & 0.03897486  & 90.6101       & 0.0154544         & 13.2 \\ 
M2V     & NGTS J062005.7-372555 & 5575203489668007936 & 393481864 & 95.023721  & -37.431899 & 0.400268 & 0.407808 & 0.02305492  & 87.7167       & 0                 & 13.83 \\ 
M2.5V   & NGTS J045008.8-362401 & 4818804257863710336 & 77369893  & 72.536742  & -36.400372 & 0.545633 & 0.550421 & 0.05525498  & 118.868       & 0.0494539         & 13.07 \\ 
M2.5V   & NGTS J050359.5-305327 & 4876285488813663232 & 1439071   & 74.755373  & -30.999599 & 0.456069 & 0.458954 & 0.02748212  & 54.6863       & 0.00618174        & 12.8 \\ 
M2.5V   & NGTS J045901.2-305958 & 4875598534564509312 & 1535810   & 75.997888  & -30.890807 & 0.380031 & 0.389898 & 0.01965184  & 68.7496       & 0.0710901         & 13.65 \\ 
M2.5V   & NGTS J050810.8-371850 & 4823535318959536256 & 14084620  & 77.044884  & -37.313753 & 0.514811 & 0.517217 & 0.03755042  & 59.1582       & 0.0309087         & 12.29 \\ 
M2.5V   & NGTS J061516.4-360818 & 2885223381503521536 & 267327257 & 93.818249  & -36.138454 & 0.496041 & 0.497965 & 0.03437556 & 118.666        & 0.0278178         & 13.99 \\ 
M3V     & NGTS J035219.1-311459 & 4886786408973741568 & 166804322 & 58.079759  & -31.249846 & 0.544884 & 0.549588 & 0.03602405  & 123.566       & 0.0216361         & 13.63 \\ 
M3V     & NGTS J050230.0-355301 & 4824660428592359552 & 13982951  & 75.624229  & -35.883643 & 0.308218 & 0.327511 & 0.01185441  & 41.1162       & 0.00927262        & 13.1 \\ 
M3V     & NGTS J051925.5-235535 & 2958246827787260032 & 30961390  & 79.856241  & -23.926311 & 0.564934 & 0.572365 & 0.04099549  & 131.705       & 0.0865444         & 13.94 \\ 
M3V     & NGTS J052116.1-322429 & 4826608831916384640 & 78053729  & 80.317094  & -32.357958 & 0.401901 & 0.409266 & 0.01826787  & 39.3307       & 0.0494539         & 12.52 \\ 
M3V     & NGTS J000722.8-293528 & 2320750123439437184 & 12418184  & 1.845141   & -29.5912   & 0.22872  & 0.257223 & 0.007069197 & 40.1093       & 0.0216361         & 13.44 \\ 
M4V     & NGTS J035624.7-311140 & 4886831592030178944 & 166869904 & 59.102836  & -31.194413 & 0.402508 & 0.409808 & 0.01474621  & 37.7653       & 0.10509           & 12.7 \\ 
M4V     & NGTS J044312.0-322643 & 4874430475258301184 & 170882537 & 70.800028  & -32.445142 & 0.512283 & 0.514585 & 0.02822354  &               & 0.0834535         & 12.6 \\ 
M4V     & NGTS J045519.0-321222 & 4873878176823736192 & 1357792   & 73.829291  & -32.206129 & 0.170431 & 0.201449 & 0.003005163 & 21.8796       & 0.0865444         & 13.72 \\ 
M4V     & NGTS J050423.8-373021 & 4823476460727785728 & 14001734  & 76.099199  & -37.505698 & 0.434448 & 0.438763 & 0.02066878  &               & 0.148362          & 12.42 \\ 
M4V     & NGTS J2341092-363819  & 2311548448064869120 & 224276435 & 355.288136 & -36.638609 & 0.253768 & 0.279818 & 0.005977345 & 38.7917       & 0.0649083         & 13.79 \\ 
\bottomrule

\end{tabular}}
\caption{The Spectral type, NGTS identifier, Gaia source ID, Tess Input Catalog (TIC) ID, RA, Dec, Stellar Mass (in Solar mass units), Stellar Radius (in Solar radi units), Distance (in parsecs), Stellar Luminosity (in Solar luminosity units), and the Macroscopic Flare Rate (per hour) for the stars used in the analysis.  The Stellar masses, radi. and luminosity data is from the Tess Input Catalog release  V8.  \citep{Stassun:2018} }
\label{tab:StellarParam}
\end{sidewaystable*}
\clearpage

\section{Re-Calculated Fourier Simulation Heat Map}
\begin{figure*}[!h]
  \centering
  \includegraphics[ clip=true, width=\textwidth, angle=0]{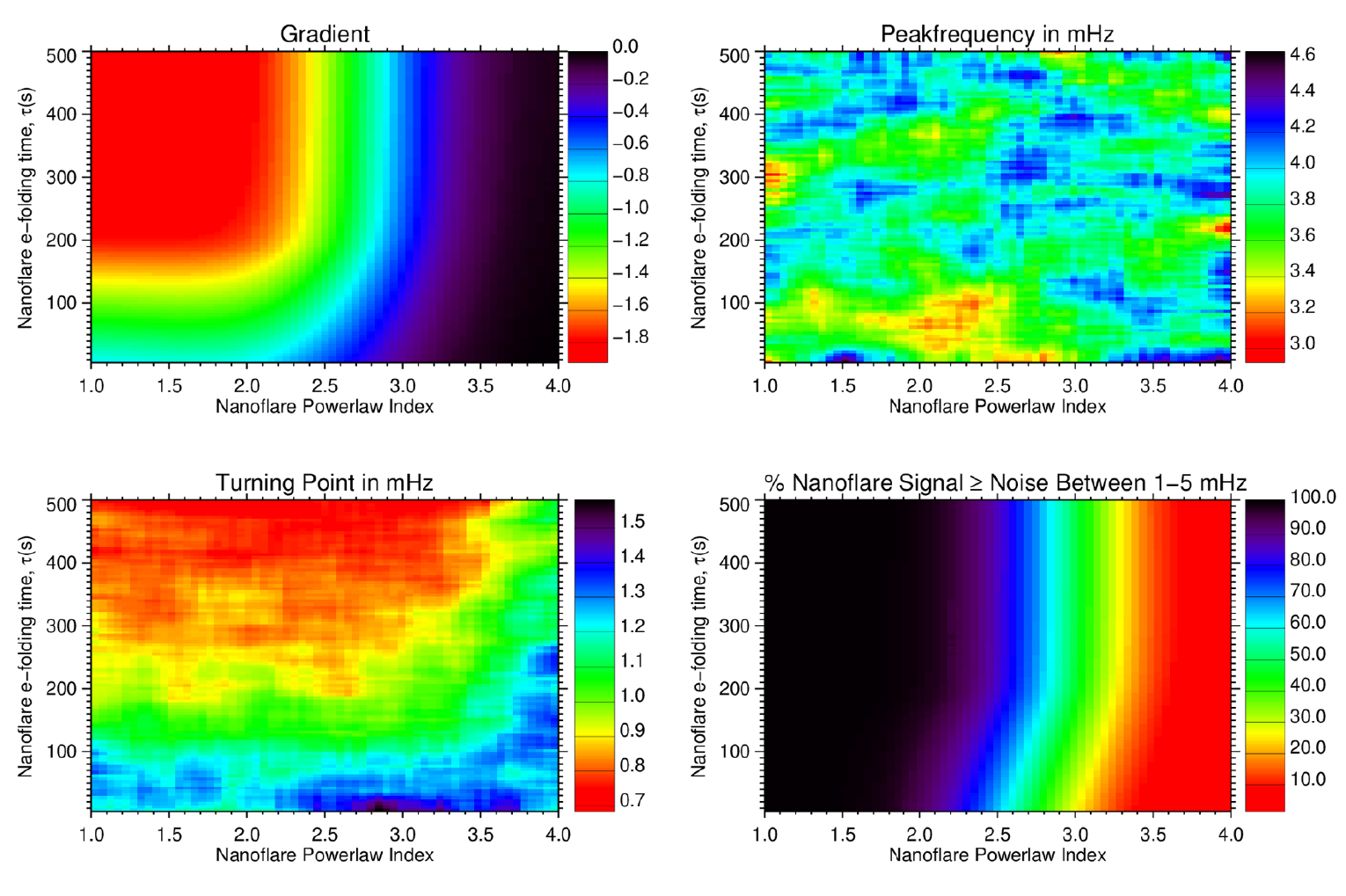}
      \caption{A reproduction of Figure 7 from \citetalias{Dillon:2020}, with the constituent PSDs re-calculated for $2095$ datapoints to match the longest continuous time series used in the present study. The primary peak frequencies (lower-left), spectral slopes (upper-left), dominant frequencies following detrending (upper-right), and the percentage of nanoflare power above the noise floor in the range of $1-5$~mHz (lower-right), is displayed as a function of the power-law index, $\alpha$, and the decay timescale, $\tau$, used to generate the synthetic time series. While a few individual values differ, the overall trends and the magnitude of the derived signals are consistent with the PSD properties generated from $2316$ datapoints and reported by \citetalias{Dillon:2020}. }
         \label{HeatMapReGen}
  \end{figure*}

Figure~{\ref{HeatMapReGen}} shows a `heat map' of the simulated PSDs \citep[c.f., Figure 7 of][]{Dillon:2020}, which has been recalculated for the $2095$ element long time series employed in the present study. While the frequency resolution is slightly reduced ($\Delta{f} = 0.0398$~mHz) than that utilized by \citet[][$\Delta{f} = 0.0356$~mHz]{Dillon:2020}, the overall trends and evolution remain consistent across the power-law index and $e$-folding timescale values.  
\label{sec:Appendix_B}

\clearpage

\bibliography{bib.bib}{}
\bibliographystyle{aasjournal.bst}



\end{document}